\documentclass[aps,prx,10pt,twocolumn,amsmath,amssymb,nofootinbib,superscriptaddress]{revtex4-2}
\usepackage{graphics,amssymb,amsmath,epsfig,color}
\usepackage{graphicx}
\usepackage[dvipsnames]{xcolor}
\usepackage{dcolumn}
\usepackage{bm}
\usepackage[colorlinks=true,citecolor=cyan]{hyperref}
\hypersetup{colorlinks=true,citecolor=cyan,linkcolor=red,urlcolor=magenta}
\usepackage{braket}
\usepackage[normalem]{ulem}
\usepackage{cancel}
\usepackage{xfrac}
\usepackage{amsfonts}
\usepackage{amsmath}
\usepackage{mathtools}
\usepackage{amssymb}
\usepackage{physics}
\usepackage{svg}
\usepackage{ORCIDinREVTeX}
\usepackage{lipsum}
\usepackage{thmtools}
\usepackage{thm-restate}
\usepackage{float}

\usepackage[capitalise]{cleveref}
\crefname{section}{Sec.}{Secs.}
\crefname{appendix}{App.}{Apps.}
\crefname{table}{Tab.}{Tabs.}
\crefname{figure}{Fig.}{Figs.}
\crefname{definition}{Def.}{Defs.}
\crefname{lemma}{Lem.}{Lems.}
\crefname{theorem}{Thm.}{Thms.}
\crefname{corollary}{Cor.}{Cors.}
\crefname{adefinition}{Def.}{Defs.}
\crefname{alemma}{Lem.}{Lems.}
\crefname{atheorem}{Thm.}{Thms.}
\crefname{acorollary}{Cor.}{Cors.}

\newtheorem{theorem}{Theorem}
\newtheorem{atheorem}{Theorem}[section]

\newtheorem{acorollary}{Corollary}[section]
\newtheorem{definition}{Definition}
\newtheorem{adefinition}{Definition}[section]
\newtheorem{lemma}{Lemma}
\newtheorem{alemma}{Lemma}[section]

\newcommand{\galg}{\mathfrak{g}}
\newcommand{\kalg}{\mathfrak{k}}
\newcommand{\malg}{\mathfrak{m}}
\newcommand{\halg}{\mathfrak{h}}
\newcommand{\balg}{\mathfrak{b}}
\newcommand{\ham}{H}

\begin{document}

\title{
RedCarD: 
A Quantum Assisted Algorithm for Fixed-Depth Unitary Synthesis \newline{}via Cartan Decomposition}

\author{Omar Alsheikh}
\orcid{0009-0008-2012-5085}
\email{ooalshei@ncsu.edu}
\affiliation{Department of Physics and Astronomy, North Carolina State University, Raleigh, North Carolina 27695, USA}

\author{Efekan K\"okc\"u}
\orcid{0000-0002-7323-7274}
\affiliation{Applied Mathematics and Computational Research Division,
            Lawrence Berkeley National Laboratory,
            Berkeley, CA 94720, USA}

\author{Bojko N. Bakalov}
\orcid{0000-0003-4630-6120}
\affiliation{Department of Mathematics, North Carolina State University, Raleigh, North Carolina 27695, USA}

\author{Alexander F. Kemper}
\orcid{0000-0002-5426-5181}
\email{akemper@ncsu.edu}
\affiliation{Department of Physics and Astronomy, North Carolina State University, Raleigh, North Carolina 27695, USA}

\date{\today}

\begin{abstract}
A critical step in developing circuits for quantum simulation is to synthesize a desired unitary operator using the circuit building blocks.
Studying unitaries and their generators from the Lie algebraic perspective has given rise to several algorithms for synthesis based on a Cartan decomposition of the dynamical Lie algebra.
For unitaries of the form $e^{-itH}$, such as time-independent Hamiltonian simulation, the resulting circuits have depth that does not depend on simulation time $t$. 
However, finding such circuits has a large classical overhead in the cost function evaluation and the high dimensional optimization problem.
In this work, by further partitioning the dynamical Lie algebra, we break down the optimization problem into smaller independent subproblems. 
Moreover, the resulting algebraic structure allows us to easily shift the evaluation of the cost function to the quantum computer, further cutting the classical overhead of the algorithm. 
As an application of the new hybrid algorithm, we synthesize the time evolution unitary for the 4-site transverse field Ising model on several IBM devices and Quantinuum's H1-1 quantum computer.
\end{abstract}

\maketitle
\section{Introduction}

Unitary synthesis is a cornerstone of quantum computing, with applications ranging from state preparation \cite{cooper2010benchmark,lee2018generalized} and time evolution \cite{bauer2020quantum,bassman2021simulating,lloyd1996universal,abrams1997simulation,zalka1998simulating,jordan2012quantum} to quantum arithmetic. In a typical unitary synthesis problem, one aims to decompose a  unitary operator into a set of specific one- and two-qubit operations that allow for direct implementation on a digital quantum computer. Among the popular techniques are series expansions \cite{berry2015simulating}, product formulas \cite{lloyd1996universal,haah2021quantum,Childs2021theory}, and learning approaches \cite{cirstoiu2020variational,mhiri2025unifying}.

The study of such unitaries and their generators from a group theoretic and a Lie algebraic perspective has been a topic of recent interest \cite{kokcu2022fixed,kokcu2024classification,wiersema2024classification,aguilar2024full,wierichs2025recursive}. A particularly useful tool is the \emph{Cartan decomposition} (CD), which was used to find the optimal circuit structure for an arbitrary two-qubit unitary \cite{vidal2004universal}. Algorithms for more general unitaries based on the CD have also been developed, often in the context of time evolution where one typically finds exponentially deep circuits for non-fast-forwardable models \cite{khaneja2005constructive,earp2005constructive,drury2008quantum_shannon,daugli2008general}.
A general unitary acting on an $n$-qubit system belongs to the special unitary group $\mathrm{SU}(2^n)$, and different CDs can be found on the group level. For unitaries with a known generator, \textit{i.e.}, those that can be expressed as $e^{A}$, where $A$ is an element of the Lie algebra $\mathfrak{su}(2^n)$, CDs can also be studied on the Lie algebraic level. A quintessential example for this case is the unitary for time evolution under a time-independent Hamiltonian $\ham$, which takes the form $U(t) = e^{-itH}$.
\begin{figure}[t]
    \centering
    \includegraphics[width=\columnwidth, clip=true, trim=0 0 0 0]{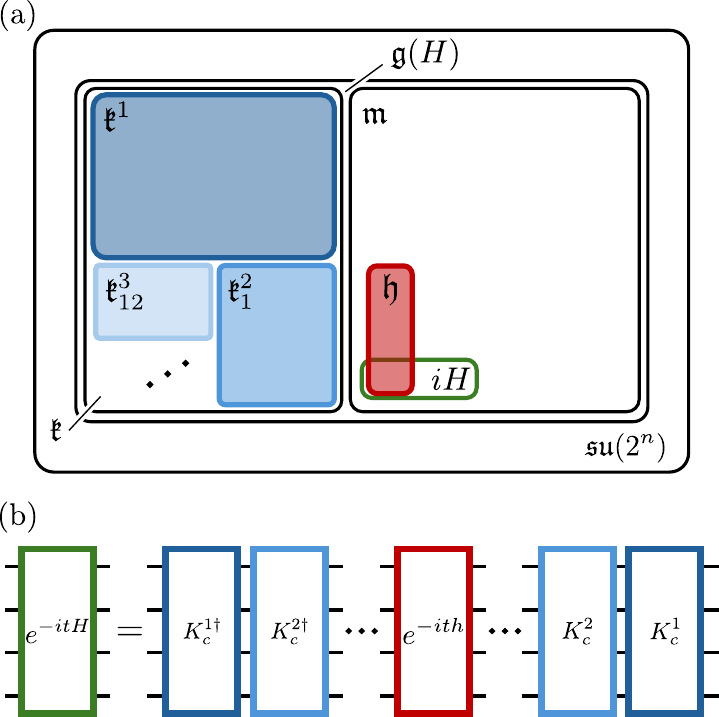}
    \caption{(a) Schematic for the Cartan decomposition of the DLA and subsequent fragmentation of $\kalg$. Here $\kalg^r_{1\dots r-1}$ is the subspace spanned by the Pauli strings that commute with the first $r-1$ basis elements of $\halg$ and anticommute with the $r^{\text{th}}$ basis element of $\halg$ (see \cref{def:ksets}). (b) Fixed-depth circuit compilation of the unitary $e^{-it\ham}$. Each unitary block $K^r_c$ is a product of the exponentials of Pauli strings in $\kalg^r_{1\dots r-1}$ (see \cref{eq:prodkr}).}
    \label{fig:newdecomp}
\end{figure}

In Refs.~\onlinecite{kokcu2022fixed,wierichs2025recursive,khaneja2005constructive,earp2005constructive,drury2008quantum_shannon,daugli2008general}, the CD was deployed to generate fixed depth circuits for $U(t)$, i.e., with a depth that does not depend on simulation time $t$. The algorithm of Ref.~\onlinecite{kokcu2022fixed} consists of three main steps: The first step is to construct the \emph{dynamical Lie algebra} (DLA) $\galg(\ham) \subseteq \mathfrak{su}(2^n)$ generated by the Pauli strings in $\ham$. This is done by finding the closure under commutation of these Pauli strings. The second step is finding a CD of $\galg(\ham) = \kalg \oplus \malg $ such that $i\ham$ is in $\malg$ (see \cref{fig:newdecomp}(a)). The third step is to classically optimize a cost function to finally find a fixed-depth circuit decomposition for $U(t)$. For fast-forwardable models \cite{gu2021fast}, the circuit depth, and thus the number of parameters to optimize over, scales polynomially in system size $n$. More typically \cite{kokcu2024classification,wiersema2024classification,aguilar2024full}, the dimension of $\galg(\ham)$ grows exponentially with system size, leading to an exponential number of parameters to be optimized. This fact has throttled the use of the algorithm by the classical optimization step, even if a fault-tolerant quantum device is at hand to perform the evolution.

In loose terms, the algorithm aims to ``rotate'' $U(t)$ into an Abelian subgroup of $\exp(\galg)$ called the Cartan subgroup (CSG). This is done by rotating $\ham$ into the generator space of the CSG, or the Cartan subalgebra $\halg$: $\ham$ is transformed into some element $h$ such that $ih \in \halg$ that commutes with a $v \in \halg$ whose exponential map $e^{sv}$ for $s\in \mathbb{R}$ defines a space-filling path that covers the entire CSG.
One then needs to optimize over as many angles as the dimension of $\kalg$.

In this work, we further leverage the algebraic structure of $\galg(\ham)$ and its CD to break down the transformation of $\ham \to h$ into intermediate steps. 
The procedure primarily relies on the observation that the basis elements of $\kalg$ can be grouped by whether they commute with the basis elements of $\halg$. For any choice of ordering of basis elements, the first set should not commute with the first element, the second set is those that commute with the first but not the second, etc. (see \cref{fig:newdecomp}(a)). As we formally show, these sets can be optimized independently, and each subsequent set is either the same size or smaller than the previous one.
Thus, instead of rotating $\ham$ into the Abelian subalgebra $\halg$ in one go, we can sequentially transform it so that at every step we get rid of a piece that lies outside of $\halg$. To do so, at every step, we need to optimize over only a fraction of the original angles. This leads to a significant reduction in the classical overhead of the algorithm, and a composite circuit structure for evolution as shown in \cref{fig:newdecomp}(b).

In addition to breaking down the original optimization problem into smaller subproblems, the new improvement allows us to completely perform all cost function calculations on quantum hardware, further reducing the classical cost to the generation of $\galg(\ham)$ and its CD and turning the algorithm into a hybrid quantum-classical algorithm called Reductive Cartan Decomposition (RedCarD).

This paper is organized as follows. In \cref{sec:recap}, we recall the definitions and results of Ref.~\onlinecite{kokcu2022fixed}. We then develop the algorithmic improvements, highlighting the decrease in classical cost. In \cref{sec:QAD}, we show how the optimization can be done on quantum hardware and provide results for a full optimization routine for a 4-site transverse field Ising model (TFIM) run on several IBM devices as well as Quantinuum's H1-1 device. In \cref{sec:discussion}, we provide a brief discussion and future directions.

\section{Unitary Synthesis via Cartan Decomposition}\label{sec:recap}
Throughout this paper, we denote vector spaces and Lie algebras by fraktur typeface $\kalg, \malg, \halg$. We denote elements by $k,m,h$. We denote the sets of $i$ times basis elements by $\tilde{\kalg},\tilde{\malg},\tilde{\halg}$, and $i$ times the basis elements by $\tilde{k},\tilde{m},\tilde{h}$. We also restrict the bases to single Pauli strings multiplied by $i$. Finally, we denote Lie groups by calligraphic typeface $\mathcal{K},\mathcal{M}$, and the elements of such groups by their roman counterparts ($K,M$).
To connect these definitions, consider a set of basis elements $\{i\tilde{k}_j\}$ for the algebra $\kalg$.
We then have $\kalg = \mathrm{span}_{i\mathbb{R}}\:\tilde{\kalg} = \mathrm{span}_{i\mathbb{R}}\{\tilde{k}_j\}$, and $\mathcal{K} = \exp (\kalg)$. Note that $|\tilde{\kalg}| = \dim\kalg = \dim \mathcal{K}$.

\subsection*{Background: Cartan decomposition}
We briefly review the original algorithm. Given a Pauli Hamiltonian $\ham$, we construct the DLA $\galg(\ham)$ (termed the \textit{Hamiltonian algebra} in Ref.~\onlinecite{kokcu2022fixed})  by finding the closure under commutation of the set of Pauli strings in $\ham$. Then, we find a Cartan decomposition (CD) $\galg(\ham) = \kalg \oplus \malg$ where
\begin{align}
    [\kalg,\kalg] \subseteq \kalg, \qquad [\malg, \malg] \subseteq \kalg,\qquad [\kalg,\malg]\subseteq \malg,
\end{align}
and $i\ham \in \malg$, and pick a maximal Abelian subalgebra --- a Cartan subalgebra --- $\halg \subseteq \malg$.
We now restate the theorem from Ref.~\onlinecite{kokcu2022fixed}, modified from Ref.~\onlinecite{earp2005constructive}:
\begin{theorem}[Improved KHK Decomposition]\label{thm:earppachos}
Assume a set of coordinates $\Vec{\theta}$ in a chart of the Lie group $\mathcal{K} = \exp(\kalg)$.
For $i\ham \in \malg$ and $K(\vec{\theta}) \in \mathcal{K}$, define the following function:
\begin{align}\label{eq:Earp_function}
    f(\vec{\theta}) = i\langle K(\Vec{\theta}) v K(\Vec{\theta})^\dagger, \ham \rangle,
\end{align}
where $\langle .,. \rangle$ denotes an invariant non-degenerate bilinear form on $\galg$, and $v \in \halg$ is an element whose exponential map $e^{sv}$ for $s\in \mathbb{R}$ is dense in $\exp(\halg)$. Then for any local extremum of $f(\Vec{\theta})$ denoted by $\Vec{\theta}_c$, and defining the critical group element $K_c = K(\Vec{\theta}_c)$, we have
\begin{align}
    ih := iK(\Vec{\theta}_c)^\dagger \ham K(\Vec{\theta}_c)= iK_c^\dagger \ham K_c  \in \halg.
    \label{eq:critical_k}
\end{align}
\end{theorem}
In other words, by finding any extremum of the cost function $f(\vec{\theta})$, we can find a representation of the unitary for time evolution $\mathcal{U}(t) = K_ce^{-iht} K_c^\dagger$, and $t$ is now a parameter that only appears in the Abelian section of the circuit. 
The density requirement for $e^{sv}$ can be easily achieved by
choosing $v = \sum \gamma_j \tilde{h}_j$, where $\{\gamma_j\}$ is a set of mutually irrational numbers.
As for $K(\vec{\theta})$, since we only need a local extremum, we do not need to fully cover $\mathcal{K}$, but rather we need a chart that contains at least one critical point. It was shown in Ref.~\onlinecite{kokcu2022fixed} that expressing $K$ as a factorized product  
\begin{align}\label{eq:prl_prodk}
    K(\vec{\theta}) = \prod_j e^{i\theta_j \tilde{k}_j},
\end{align}
is sufficient. This expression has two main advantages. First, it can be readily integrated in a quantum circuit. Second, it offers an efficient way of calculating $KvK^\dagger$, which appears in the cost function $f(\vec{\theta})$, requiring only $\mathcal{O}(|\tilde{\kalg}||\tilde{\malg}|)$ Pauli string products. This is in contrast to using the exponential map $e^{i\sum \theta_j \tilde{k}_j}$ directly, for which the calculation scales as $\mathcal{O}(|\tilde{\kalg}|^2|\tilde{\malg}|)$.

\subsection*{Reductive Cartan Decomposition (RedCarD)}

The bottleneck of the algorithm is the classical optimization of $f(\vec{\theta})$. For nonintegrable models, the dimension of
$\mathfrak{g}$ increases exponentially with the number of sites $n$ \cite{wiersema2024classification,kokcu2024classification}, which translates to an exponential number of parameters to optimize. 
In addition to this being a large optimization problem, the
cost function can be expensive to evaluate.
Here, we present a procedure that breaks down the optimization problem into smaller subproblems with decreasing size by grouping the basis elements of $\kalg$ based on whether or not they commute with the sequence of basis elements of $\halg$.
We start by defining the following structure:
\begin{definition}\label{def:symmetric_r}
    For any two vector spaces $\mathfrak{a} = \:\mathrm{span}_{i\mathbb{R}} \{ \tilde{a}_s\}$ and $\balg = \:\mathrm{span}_{i\mathbb{R}} \{ \tilde{b}_j \}$ where $\tilde{a}_s$ and $\tilde{b}_j$ are Pauli strings, define the $r$-symmetric subspace of $\mathfrak{a}$ as 
    \begin{align}
        \mathfrak{a}_{(r)}(\balg) := \mathrm{span}_{i\mathbb{R}} \{ \tilde{a}_s \:\:|\:\: [\tilde{a}_s,\tilde{b}_j] = 0 \ \forall\ j<r \}.
    \end{align}
\end{definition}
These subspaces allow us to further decompose the CD of $\galg(\ham)$ through the following lemma:
\begin{lemma}[Reductive Cartan Decomposition]\label{lem:reductivecd}
    Given a Cartan decomposition $\galg = \kalg \oplus \malg$ and a vector space $\balg = \mathrm{span}_{i\mathbb{R}}\{\tilde{b}_j\}$, and the $r$-symmetric subspaces $\galg_{(r)}(\balg),\kalg_{(r)}(\balg),\malg_{(r)}(\balg)$, then 
    \begin{align}
    \galg_{(r)}(\balg) = \kalg_{(r)}(\balg) \oplus \malg_{(r)}(\balg)
    \end{align} 
    is a Cartan decomposition, \textit{i.e.},
    \begin{align}
    [\kalg_{(r)}(\balg), \kalg_{(r)}(\balg)] &\subseteq \kalg_{(r)}(\balg), \nonumber \\
    [\malg_{(r)}(\balg), \malg_{(r)}(\balg)] &\subseteq \kalg_{(r)}(\balg), \nonumber \\
    [\kalg_{(r)}(\balg), \malg_{(r)}(\balg)] &\subseteq \malg_{(r)}(\balg).
    \end{align}
\end{lemma}
\textbf{Proof} Consider $m_1,m_2 \in \mathfrak{m}_{(r)}(\balg)$, i.e., $[m_1,\tilde{b}_j]=[m_2,\tilde{b}_j]=0$ for all $j<r$. Thus, $\big[[m_1,m_2],\tilde{b}_j\big]=0$. Since $[m_1,m_2]\in \mathfrak{k}$, it follows that $[m_1,m_2]\in \mathfrak{k}_{(r)}(\balg)$. The other two commutation relations can be proved similarly.\hfill$\blacksquare$

This will be used for proving the next theorem. We make another definition for notational convenience.

\begin{definition}\label{def:ksets}
    Given two Pauli vector spaces $\kalg=\mathrm{span}_{i\mathbb{R}} \{\tilde{k}_j\}$ and $\balg = \mathrm{span}_{i\mathbb{R}}\{\tilde{b}_j\}$, define the following family of subspaces:
    \begin{align}
        \kalg^{i_1i_2\dots}_{j_1j_2\dots}(\balg) = \mathrm{span}_{i\mathbb{R}}\big\{\tilde{k}_j \ | \ [\tilde{k}_j,\tilde{b}_{i_r}]\neq0,\ [\tilde{k}_j,\tilde{b}_{j_s}]=0\ \forall \ r,s\big\}.
    \end{align}
\end{definition}
In the following theorem, we will be using $\kalg^{r}_{1\dots r-1}(\balg)$, which is spanned by the terms that commute with $\tilde{b}_1, \dots, \tilde{b}_{r-1}$ and anticommute with $\tilde{b}_r$. Note that $\mathrm{dim}\:\kalg^{r}_{1\dots r-1} = |\tilde{\kalg}^r_{1\dots r-1}|=|\tilde{\kalg}_{(r)}|-|\tilde{\kalg}_{(r+1)}|$.
Now we state the main theorem of the paper.

\begin{theorem}[Reductive KHK Decomposition]\label{thm:reductive_KHK}
    Given a vector space $\mathfrak{b} = \mathrm{span}_{i\mathbb{R}}\{\tilde{b}_j\} \subseteq \malg$ and the $r$-symmetric subspaces $\galg_{(r)}(\balg) = \kalg_{(r)}(\balg) \oplus \malg_{(r)}(\balg)$, take $i\ham_r \in \malg_{(r)}(\balg)$ and
    \begin{align}\label{eq:prodkr}
        K^r(\vec{\alpha}) = \prod_{\tilde{k}_j \in \tilde{\kalg}^r_{1\dots r-1}(\balg)} e^{i \alpha_j \tilde{k}_j},
    \end{align}
    and define the following function:
    \begin{align}\label{eq:new_cost}
         f_r(\vec{\alpha}) = \langle K^r(\vec{\alpha}) \tilde{b}_r K^r(\vec{\alpha})^\dagger, \ham_r \rangle.
    \end{align}
    If $\mathfrak{b}$ is Abelian, then for any \emph{local extremum} of this function denoted as $\vec{\alpha}_c$, and defining the critical group element $K_c^r = K^r(\vec{\alpha}_c)$, we have
    \begin{align}\label{eq:redcom}
        [K^{r\dagger}_c \ham_r K^r_c, \tilde{b}_j] = 0 
    \end{align}
    for all $j < r+1$, i.e., $iK^{r\dagger}_c  \ham_r K^r_c \in \malg_{(r+1)}(\balg)$.
\end{theorem}
We present a proof of the theorem in \cref{sec:reductive_KHK}.

The most important aspect of \cref{thm:reductive_KHK} is that it reduces the number of parameters we optimize over at any given step.
Let us see what happens when we apply \cref{thm:reductive_KHK} repeatedly to $\ham$. Taking $b_0 = 0$, then according to \cref{def:symmetric_r}, $\kalg_{(1)} = \kalg$ and $\malg_{(1)}=\malg$. Working with \cref{eq:prodkr}, we optimize over the 
$|\tilde{\kalg}^1
|=|\tilde{\kalg}_{(1)}|-|\tilde{\kalg}_{(2)}|$ terms that do not commute with $b_1$ to obtain $i\ham_2 = iK^{1\dagger}_c \ham K^1_c \in \malg_{(2)}$. After $r$ optimizations, we arrive at $i\ham_{r+1} = iK^{r\dagger}_c \cdots K^{1\dagger}_c \ham K^1_c \cdots K^r_c \in \malg_{(r+1)}$. If we choose $\balg = \halg$, then $i\ham_{|\tilde{\halg}|+1} \in \halg$, which is the desired outcome of \cref{thm:earppachos} and fulfills the goal of synthesizing the unitary. In this way, \cref{thm:reductive_KHK} allows us to transform the Hamiltonian in a series of optimization procedures over $|\tilde{\kalg}^{1}|$, $|\tilde{\kalg}^{2}_1|$, $\cdots$, $\left|\tilde{\kalg}^{|\tilde{\halg}|}_{1\dots |\tilde{\halg}|-1}\right|$ parameters. Considering that the time and number of operations required in an optimization routine for convergence increase worse than linearly as the number of parameters grows, the reductive KHK decomposition gives a huge improvement over the original algorithm. 

An important remark here is that we are not restricted to an ansatz of the form \cref{eq:prodkr} to perform a reductive KHK decomposition. As we discussed before, our choice of this ansatz stems from its readiness to be compiled into a quantum circuit and the lower cost of cost function evaluations. However, any coordinate patch of the group $\mathcal{K}_{(r)} = \exp(\kalg_{(r)})$ would lead to a reductive KHK decomposition. 
At first glance, this would mean that we have to solve a $\big|\tilde{\kalg}_{(r)}\big|$-dimensional optimization problem at the $r^{\text{th}}$ step,
but we can reduce the number of optimization parameters to $\big|\tilde{\kalg}_{(r)}\big| - \big|\tilde{\kalg}_{(r+1)}\big|$. To do so,
we take a coordinate patch of the form $K^r(\vec{\alpha})S^r(\vec{\beta})$, where $S^r(\vec{\beta})$
is itself a coordinate patch of the subgroup $\mathcal{K}_{(r+1)} = \exp(\kalg_{(r+1)}) \subseteq \mathcal{K}_{(r)}$, \textit{i.e.}, it commutes with $\tilde{h}_r$
(see \cref{sec:reductive_KHK}).
Thus, we only need to optimize over $K^r(\vec{\alpha})$, and the next steps of the procedure will cover the $S^r(\vec{\beta})$ optimization.
The observation that we are not bound to the factorized ansatz of \cref{eq:prodkr} permits us to choose other ans\"atze $K^r(\vec{\alpha})$ that may be more useful in different contexts other than the one considered here. 

\begin{figure}[t]
    \centering
    \includegraphics[width=0.9\columnwidth, clip=true, trim=0 0 0 0]{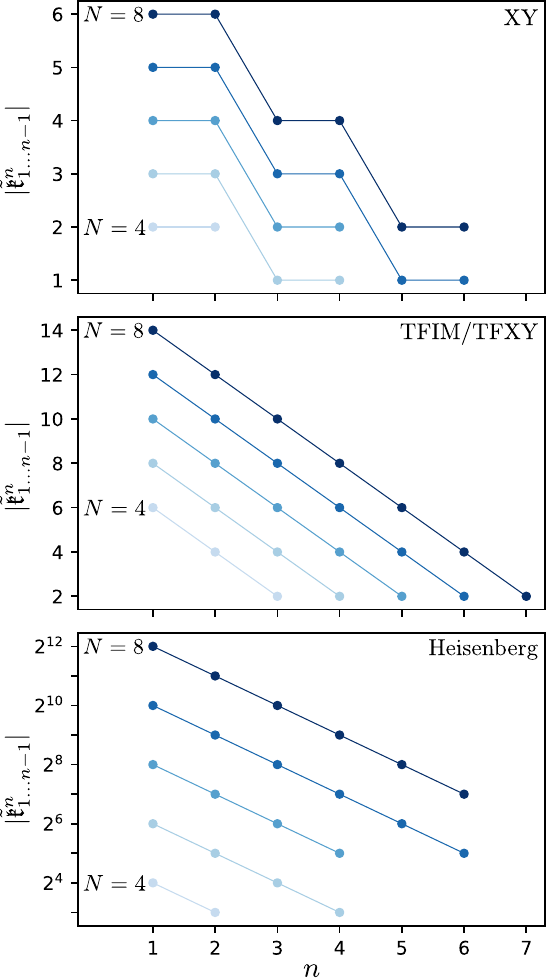}
    \caption{Number of parameters to optimize over for each step of the reductive KHK decomposition for the 1-dimensional nearest neighbor XY, transverse field Ising/XY, and Heisenberg spin models.}
    \label{fig:modelsubspaces}
\end{figure}

The algorithm suggests that we need to solve $\mathcal{O}(|\tilde{\halg}|)$ optimization problems. However, while grouping the elements in $\tilde{\kalg}$, some sets may turn out to be empty, i.e., all the remaining elements commute with the $h_r$ element at hand and the procedure terminates. In fact, for any DLA $\galg \subseteq \mathfrak{su}(2^n)$, one can show that we can have at most $n$ non-empty sets. We refer the reader to \cref{sec:associative} for a formal treatment.

\begin{figure*}[t!]
    \centering
    \includegraphics[clip=true, trim=0 0 0 0, width=0.85\textwidth]{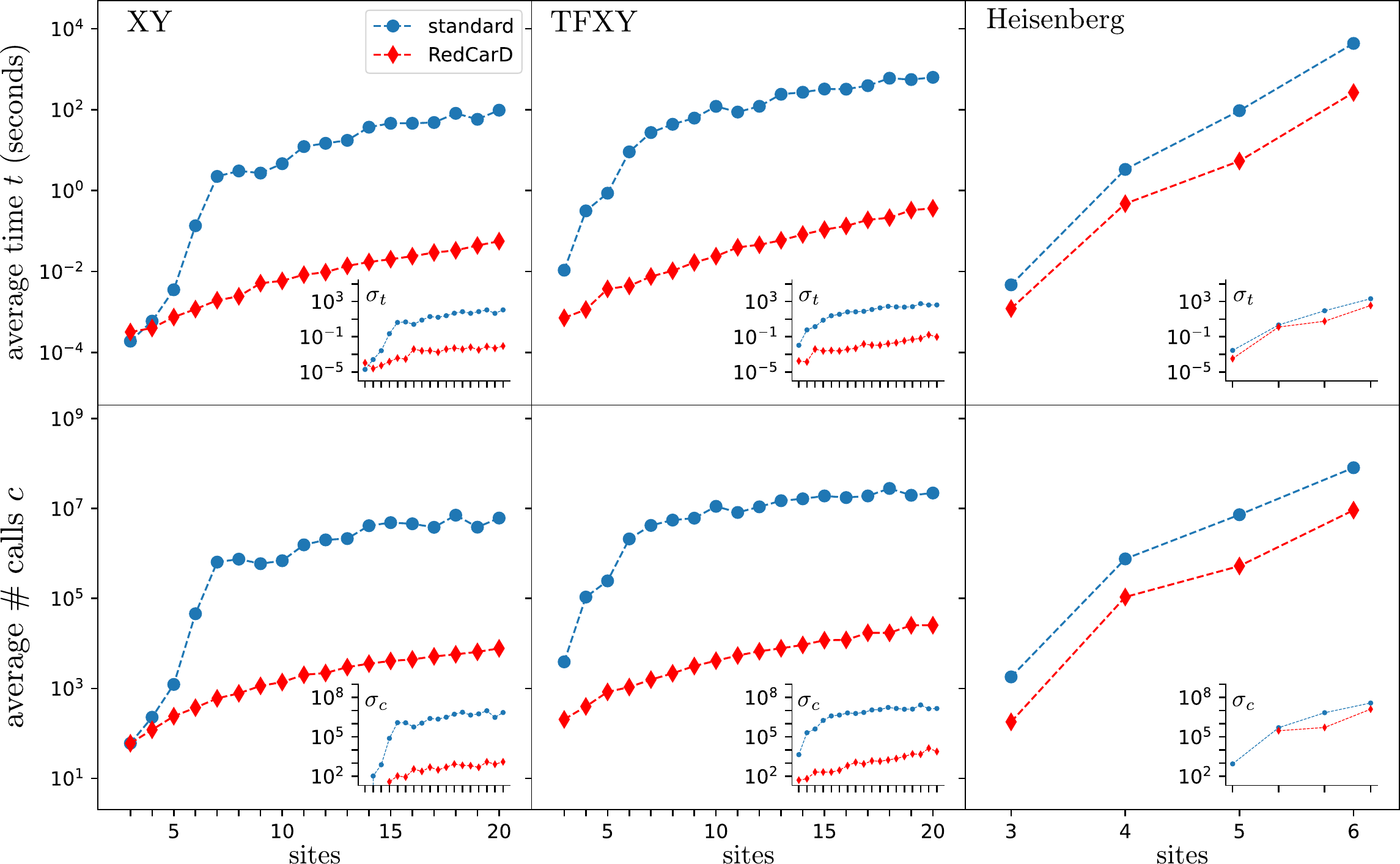}
    \caption{Runtime and cost function calls comparison between the standard algorithm and the modified version. Statistics are done on 10 converging runs. A run converges if the ratio of the Hilbert-Schmidt norm of the part of $K^\dagger \ham K$ that lies outside of $\halg$ to the Hilbert-Schmidt norm of $\ham$ reaches 1\% in less than $10^5$ iterations. The same ansatz is used for each run, and initial angles are randomized. We used the Rotosolve algorithm \cite{ostaszewski2021structure} to minimize the cost function.}
    \label{fig:comparison}
\end{figure*}

To illustrate how the reductive decomposition works, in \cref{fig:modelsubspaces} we show the number of angles to be optimized after each iteration for three 1-dimensional nearest-neighbor spin models.
We observe empirically that, with the exception of the XY model, both our choice of $\halg$ and the ordering of its basis elements $\{i\tilde{h}_1,i\tilde{h}_2,\dots\}$ do not affect the sizes of the non-empty sets. Permuting $\{i\tilde{h}_1,i\tilde{h}_2,\dots\}$ only affects where the empty sets appear. We also note that apart from the XY model, the non-empty sets are strictly decreasing in size. The different behavior of the XY model can be explained by the fact that the Hamiltonian can be written as a sum of two mutually commuting sets, where every term from the first set commutes with every term from the second.
Essentially, this means that the two sets can be treated as separate problems. All of these empirical observations are the result of a more general statement about the relative sizes of the subsets. 

Before stating the theorem, we present another definition:
\begin{definition}
    Given a set of Pauli strings $\mathcal{P}$, define a graph with a set of vertices $\mathcal{P}$ and edges connecting all pairs $\sigma_i,\sigma_j\in \mathcal{P}$ such that $[\sigma_i,\sigma_j]\neq 0$. This is called the \emph{frustration graph} of $\mathcal{P}$.
\end{definition}

We now state the following theorem:
\begin{restatable}{theorem}{ksizes}\label{thm:ksizes}
    Given a DLA $\galg$ and its Cartan decomposition $\galg = \kalg \oplus \malg$, take an Abelian subalgebra $\balg \subseteq \malg$  such that its Pauli string basis elements belong to the same connected component of the frustration graph of $\galg$, i.e., there exists at least one path connecting every pair. Then the basis elements of $\balg$ can be ordered in such a way that the following inequalities hold:
    \begin{align}
        |\tilde{\kalg}^1(\balg)| > |\tilde{\kalg}^2_1(\balg)| \geq |\tilde{\kalg}^3_{12}(\balg)| \geq \dots \geq |\tilde{\kalg}^n_{1\dots n-1}(\balg)|.
    \end{align}
    The inequalities are saturated only if the sets are empty, i.e., the empty sets appear at the end of the sequence. If the first $r$ sets are non-empty, the inequalities hold for any permutation of the first $r$ indices. A general permutation may cause the empty sets to appear earlier, but it does not change the ordering of the non-empty sets.
\end{restatable}
\noindent The proof of \cref{thm:ksizes} is derived in \cref{sec:ksizes}.

The partitioning of the optimization problem into a series of smaller sub-problems with decreasing size has a marked influence on the resources required.
In \cref{fig:comparison} we compare the runtime and the number of cost function calls required to converge using the original algorithm and the improved version (RedCarD) on a system equipped with an Intel i7 13700HX mobile processor. We present the average number as well as the standard deviation taken over 10 converging runs.
For both metrics, and for the three models considered, we require orders of magnitude less resources for the optimization as the problem size gets larger, resulting in a reduction of the total runtime; for cases such as the 20 site TFXY model the reduction is from hours to a fraction of a second.

In summary, we provide the following algorithm to obtain a fixed-depth circuit compilation for a given unitary $e^{-it\ham}$ based on the reductive Cartan decomposition (RedCarD):
\begin{enumerate}
    \item Generate the dynamical Lie algebra $\galg(\ham)$ and generate its Cartan decomposition $\galg(\ham) = \kalg \oplus \malg$ such that $i\ham \in \malg$. Construct a Cartan subalgebra $\halg \subseteq \malg$.
    \item For every basis element $i\tilde{h}_r \in \halg$, group the basis elements $i\tilde{k}_j \in \kalg$ such that $[\tilde{k}_j,\tilde{h}_r] \neq 0$ and $[\tilde{k}_j,\tilde{h}_s]=0$ for all $s<r$.  
    \item Choose the binary form $\langle A,B \rangle$ = $\Tr(AB)$. Let $r=1$ and define $\ham_1 = \ham$. 
    \item Find a local extremum of $f_r(\vec{\alpha})$ given in \cref{thm:reductive_KHK}. Calculate $\ham_{r+1} = K_c^{r\dagger} \ham_{r} K_c^r$.
    \item While $r<|\tilde{\halg}|$, increment $r$ by 1 and repeat step 4.
    \item Obtain $ih = iK^{|\tilde{\halg}|\dagger}_c \cdots K^{1\dagger}_c \ham K^1_c \cdots K^{|\tilde{\halg}|}_c \in \halg$. Compile the fixed-depth circuit for the unitary $U(t) = K^1_c \cdots K^{|\tilde{\halg}|}_c e^{-ith} K^{|\tilde{\halg}|\dagger}_c \cdots K^{1\dagger}_c$.
\end{enumerate}

\medskip

\subsection*{An illustrative example:\\ the transverse field Ising model}

We demonstrate how RedCarD works for a 1-dimensional transverse field Ising model (TFIM).
The Hamiltonian for $l$ sites is
\begin{align}
    \ham = -J\sum_{i=1}^l X_i X_{i+1} + g\sum_{i=1}^l Z_i.
\end{align}
We will use the following CD for this model
\cite{kokcu2022fixed}:
\begin{align}\label{eq:tfim_cd}
    \kalg &= \mathrm{span}_{i\mathbb{R}}\{\widehat{X_iY_j}, \widehat{Y_iX_j}\: |\: i,j=1,2,\dots,l,\ i<j \}, \nonumber \\
    \malg &= \mathrm{span}_{i\mathbb{R}}\{Z_k,\widehat{X_iX_j}, \widehat{Y_iY_j}\: |\: i,j,k=1,2,\dots,l,\ i<j \}, \nonumber \\
    \halg &= \mathrm{span}_{i\mathbb{R}}\{Z_k \:|\: k=1,2,\dots,l\},
\end{align}
where
\begin{align}
    \widehat{A_iB_j} := A_i Z_{i+1}Z_{i+2}\cdots Z_{j-1}B_j.
\end{align}

\begin{figure}[t]
    \centering
    \includegraphics[width=\columnwidth]{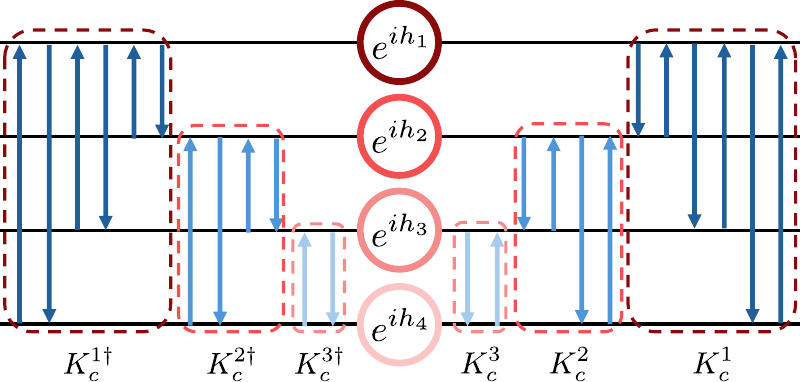}
    \caption{Circuit for a 4-site TFIM time evolution unitary. An arrow doublet connecting two qubits $p$ and $q$ depicts the unitary $e^{i\alpha \widehat{X_pY_q}+i\beta \widehat{Y_pX_q}}$. Flipping the doublet represents Hermitian conjugation. The color of a dashed box indicates the basis element of $\halg$ with which the doublets do not commute.
    }
    \label{fig:TFIMcirc}
\end{figure}

We note that for this model the basis of the Cartan subalgebra $\halg$ consists of Pauli $Z$ operators acting on the individual sites. Moreover, the basis elements of $\kalg$ are all strings of $Z$ terminated by an $X$ or $Y$. This suggests an illustrative pictorial representation, shown in \cref{fig:TFIMcirc} for 4 sites. The arrows are elements of $\kalg$, grouped by whether or not they commute with the elements of $\halg$ in increasing site order. This representation makes it clear why the reductive approach works: the elements of $\kalg$ that do not touch the $1^{\text{st}}$ qubit do not affect the part of the problem where the optimization with respect to $h_1$ is done, and so forth.

More precisely, going through the basis elements of $\halg$ one by one, starting with $Z_1$, we group the basis elements of $\kalg$ so that they do not commute with the current element of $\halg$ and commute with all the previous ones:
\begin{align}\label{eq:TFIMkspaces}
    \kalg^1 &= \mathrm{span}_{i\mathbb{R}}\{\widehat{X_1Y_j}, \widehat{Y_1X_j} \: | \: j=2,3,\dots,l\}, \nonumber \\
    \kalg^2_1 &=  \mathrm{span}_{i\mathbb{R}}\{\widehat{X_2Y_j}, \widehat{Y_2X_j} \: | \: j=3,4,\dots, l\}, \nonumber \\
    &\vdotswithin{ = } \notag \\
    \kalg^r_{1\dots r-1} &= \mathrm{span}_{i\mathbb{R}}\{\widehat{X_rY_j}, \widehat{Y_rX_j} \: | \: j=r+1,r+2,\dots, l\}, \nonumber \\
    \kalg^l_{1\dots l-1} &= \{0\}.
\end{align}
This grouping results in the circuit shown in \cref{fig:TFIMcirc}.

\section{Quantum Assisted Decomposition}\label{sec:QAD}

Not only does RedCarD give an improved exponential scaling over its standard counterpart, but it also allows for performing all the cost function calculations needed on a quantum device. Taking the trace as the binary form on the dynamical Lie algebra, we can write the cost function as
\begin{align}\label{eq:costcirc}
    \mathrm{Tr}\big(K^r(\vec{\alpha})\rho_r K^r (\vec{\alpha})^\dagger \ham_r \big)= \frac{1}{2^n}f_r(\vec{\alpha}),
\end{align}
where
\begin{align}\label{eq:rhob}
    \rho_r = \frac{I + \tilde{h}_r}{2^n},
\end{align}
since $\mathrm{Tr}(\ham_r)$ is a constant which can be ignored. 
\cref{eq:costcirc} resembles an expectation value of the transformed Hamiltonian $\langle K^{r\dagger} \ham_r K^r\rangle$ of the mixed state $\rho_r$. As we show in \cref{sec:statecircuits}, it is straightforward to prepare the mixed state \cref{eq:rhob} when $\tilde{h}_r$ is a single Pauli string; in contrast, we are not aware of an efficient way to prepare the mixed state $\rho$ when $\tilde{h}_r$ is replaced with $v$ that appears in the original algorithm.

We also note that, due to our choice of the coordinate patch in \cref{eq:prodkr}, the cost function as a function of any given $\alpha_j$ is a simple harmonic with a period $\pi$. If we therefore evaluate the cost function at 3 points for a given $\alpha_j$, we can fit it to a sinusoid and easily find its minimum. Doing this sequentially for all angles, we are guaranteed to lower the cost function after each step. This gradient-free procedure is known as Rotosolve \cite{ostaszewski2021structure}. This particular procedure is much more resistant to hardware noise compared to methods that require a gradient or other form of optimization landscape characterization.

\begin{figure}[b]
    \centering
    \includegraphics[width=\columnwidth]{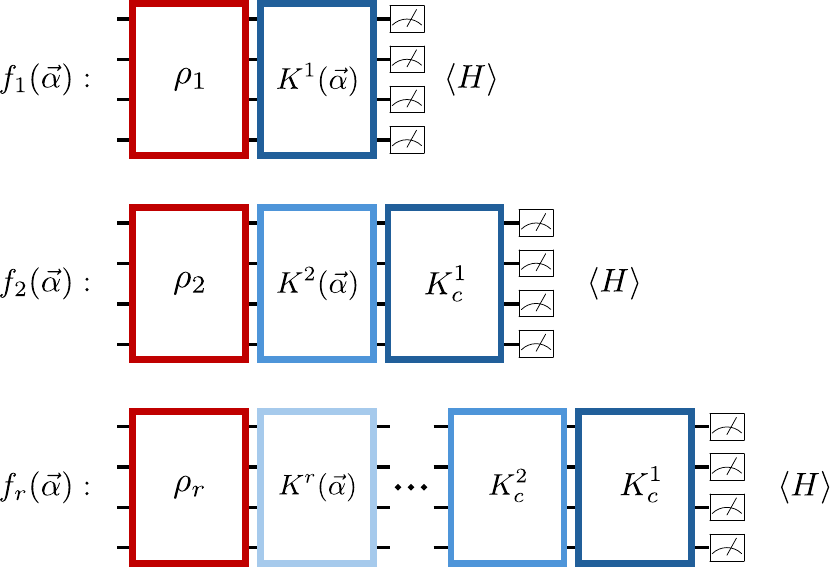}
    \caption{Quantum assisted KHK decomposition. At each step, only the angles in $K^r$ are varied as the previous angles are already fixed from the previous optimizations.}
    \label{fig:rotocircuit}
\end{figure}

\begin{figure*}[htpb]
    \centering
    \includegraphics[clip=true, trim=0 0 0 0,width=0.8\textwidth]{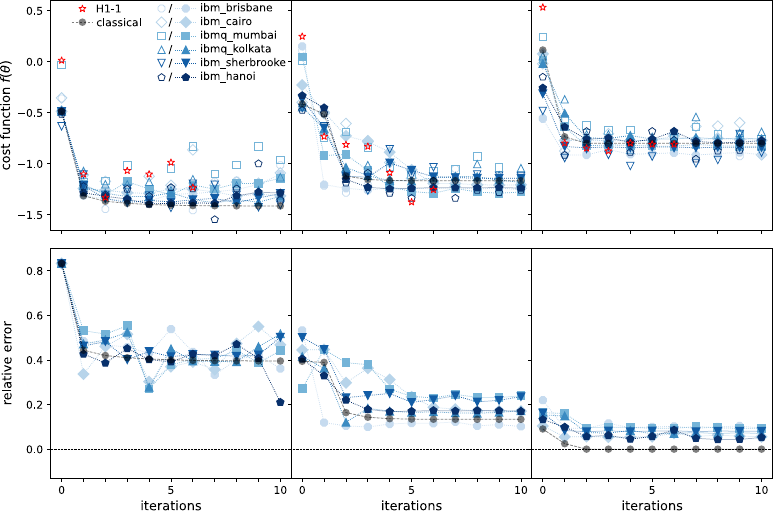}
    \caption{Quantum assisted KHK decomposition for a 4-site TFIM. The relative error is defined as the ratio of the Hilbert-Schmidt norm of the part of $K^\dagger \ham K$ that lies outside of $\halg$ to the Hilbert-Schmidt norm of $\ham$. We used 800 shots for the IBM devices and 400 shots for Quantinuum's H1-1. The IBM runs all had the same initial set of angles. For the IBM devices, solid marks denote cost function values calculated classically using the angles obtained from quantum hardware, while hollow marks are cost function evaluations on quantum hardware. H1-1 data points are cost function evaluations obtained from the quantum device using angles obtained from a classical run. Results are plotted against a full classical optimization run with the same initial angles as the IBM runs for comparison.}
    \label{fig:HW}
\end{figure*}

These observations can be combined to transfer 
the expensive evaluation of the cost function portion of the unitary synthesis procedure to quantum hardware.
\cref{fig:rotocircuit} shows the setup for the quantum assisted optimization circuits. 
For a given optimization subproblem, the cost function minimization is achieved as follows:
\begin{enumerate}
    \item Prepare the state $\rho_r$. Initialize a random vector $\vec{\alpha}$.
    \item Select an angle $\alpha_j$ to optimize, and choose 3 values $\alpha_j \in [0,\pi]$. Evolve $\rho_r$ with $K^r(\vec{\alpha})$ at these 3 points and measure $\langle \ham_r \rangle$.
    \item Fit $\langle \ham_r \rangle$ to a sinusoid. Determine the angle $\alpha_c$ that minimizes the function.
    \item Update $\alpha_j$ to $\alpha_c$.
    \item Repeat steps 2-4 until all the angles are updated. This constitutes one iteration.
    \item Perform as many iterations as needed until a stop criterion is met.
    \item Obtain $\ham_{r+1} = K_c^{r\dagger}\ham_r K_c^r$.
\end{enumerate}
The evaluation of the cost function, and thus the complexity of the algorithm as a whole, scales as $O(|\tilde{\kalg}^r_{1\dots r-1}||\tilde{\malg}_{(r)}|)$. Having the ability to calculate $f_{(r)}(\vec{\alpha})$ on a quantum computer allows us to avoid the costly evaluation of the cost function on a classical computer.

\subsection*{Transverse Field Ising Model: Quantum Hardware Results}

In order to demonstrate our algorithm, we optimized a 4-site transverse field Ising model (TFIM) on IBM devices and Quantinuum's H1-1 hardware. We set $J=1$ and $g=0.5$.
For the TFIM a further compression can be performed \cite{kokcu2022algebraic,camps2022algebraic}, resulting in a depth-$(n/2)$ circuit which we use here
(see \cref{sec:reductive_KHK}).
For the IBM runs, the procedure is fully quantum, which means that all the cost function calculations in the procedure were performed on the quantum device. For the H1-1 results, the angles were calculated classically, and for each angle update we measure the cost function on the quantum device.

We plot the evolution of the cost function in \cref{fig:HW}. Across all devices, we see an overall decreasing trend in the cost function. The fluctuations are more pronounced in the first optimization subproblem, and toward the end they die out. This may be explained by the fact that at the beginning we are optimizing 6 angles, while at the last subproblem we are optimizing over only 2 angles while keeping the other 10 angles fixed. 
The angles found after every iteration are affected by shot noise and hardware noise, and so we expect the variance to be bigger when we have more instances in which this noise can affect the output. As a result, the data are less spread out as we go from 6 to 4 to 2 angles. The hardware relative error plots plateau at values greater than that of the classical result due to the error accrued from the previous optimization subproblems.
\begin{figure}[t]
    \centering
    \includegraphics[clip=true, trim=0 0 0 0, width=\columnwidth]{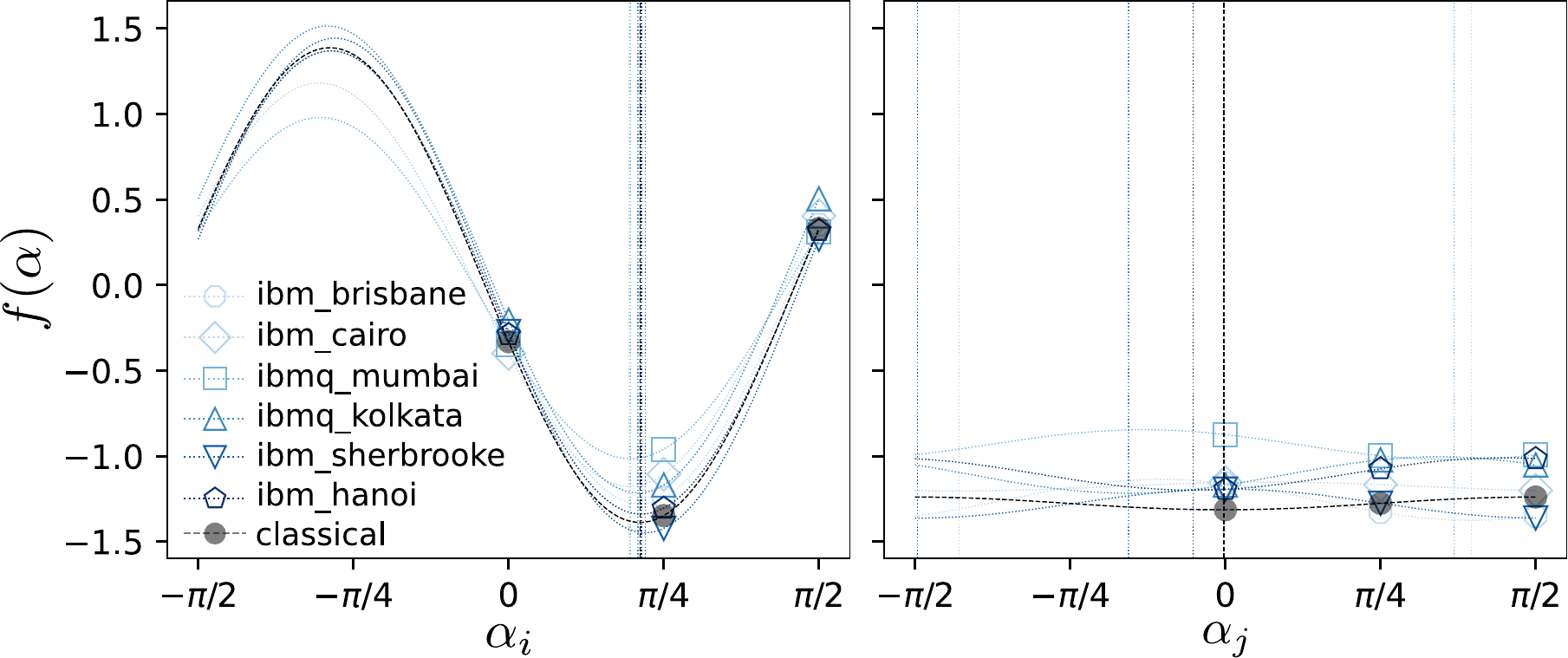}
    \caption{Two snapshots of the Rotosolve procedure. When the sinusoid amplitude is large, the minimum angle is noise robust. Conversely, when the cost function is almost flat, small fluctuations can lead to significantly different solutions.}
    \label{fig:rotoplot}
\end{figure}

Another interesting feature is that even though the error is uniformly decreasing across the different devices, the final optimized angles vary, and none of them give the angles obtained classically. This is because at some points during the procedure, the fitted sinusoid is almost flat, and this causes the results to diverge. This can be seen in \cref{fig:rotoplot} where we plot two sample iterations from the various devices. The lack of ability to resolve the optimal angles when
the function is flat does not cause any adverse effects;
rather,
it causes a shift in the trajectory of the minimization procedure toward some other region in the parameter space, and eventually toward a local minimum, which is all that is needed. 
When the amplitude of the sinusoid is small, i.e., a flat cost function for that particular angle, the optimization does not move far from the current point in the trajectory.
When the amplitude of the sinusoid is large, the minimum angle is less sensitive to noise, in particular the kinds of noise that scale expectation values such as depolarizing noise. In the end, we have a procedure that is very good at finding the minimum when the sinusoid has a large amplitude. 

\section{Discussion}\label{sec:discussion}
In this work, we presented an improved version of the algorithm proposed in Ref.~\onlinecite{kokcu2022fixed} to generate fixed-depth circuits for unitaries of the form $e^{-it\ham}$. The algorithm makes use of the algebraic structure of the Hamiltonian $\ham$ to optimize a cost function in order to generate a factorized product that can be readily compiled into a quantum circuit. We used additional properties of the algebraic structure to break down the optimization problem into smaller subproblems, which cuts down on the classical cost. The new algorithm also allows us to perform all the cost function calculations on a quantum device, which is the most expensive part of the algorithm. We demonstrated the quantum assisted algorithm on several IBM devices and Quantinuum's H1-1 hardware for a 4-site TFIM model.

The improved version still suffers from some of the drawbacks of the original algorithm. Namely, in cases where the dynamical Lie algebra (DLA) scales exponentially with system size, even though we optimize over a smaller subset of the total angles at any given time, we still need to optimize over an exponential number of them. This results in exponentially deep circuits. Despite this fact, the circuits can be used for any simulation time $t$. The algorithm can also be used to prepare states using the unitary coupled cluster (UCC) formalism \cite{cooper2010benchmark,lee2018generalized} and projection algorithms (e.g. \cite{rule2024simplified}). In general, the algorithm is useful when a fixed-depth circuit for the exponential of some algebra element is needed. The Cartan decomposition (CD) is exact; hence, the symmetries of the system are automatically preserved. This is in contrast to Trotter-based approaches, where symmetries must be explicitly accounted for when constructing circuits \cite{tran2021faster,gard2020efficient}, which can be difficult \cite{marvian2022restrictions}.

A related work that uses the CD to produce fixed-depth circuits using a non-variational approach was discussed in Ref.~\onlinecite{wierichs2025recursive}.
The algorithm decomposes a given unitary by recursively applying a CD. The algorithm exchanges the load of optimizing a non-convex function with an exponentially large linear algebra problem and a subgraph isomorphism problem. Although it benefits from having $i\ham \in \malg$, it can be applied to any Hamiltonian with any choice of a CD. In the case where $i\ham \notin \malg$, a different decomposition has to be found for every simulation time $t$. Our algorithm in its current form restricts $i\ham$ to be in $\malg$, which is sometimes impossible to do \cite{wierichs2025recursive}. Fortunately, our algorithm can be extended to cover the general case, with the advantage that we now once again only need to do the optimization once for a given $\ham$, and we get a fixed-depth decomposition valid for any $t$ \cite{GHG}. 

Currently, our algorithm relies on the full algebraic structure of the DLA $\galg(\ham)$ with no regard to the coefficients of the individual terms. Although the coefficients most likely shape how $h$ turns out, currently we have little understanding of their exact role. This is important because if one starts with a free-fermionic Hamiltonian with a polynomial scaling $\galg(\ham)$, adding a small perturbation immediately adds an exponential number of terms to the DLA. In a future direction, we hope to use the coefficients that appear in the Hamiltonian to devise a perturbative scheme of the CD, cutting down the cost of the optimization and obtaining shallower approximate circuits.

\vspace{-0.2in}

\section*{Author Contributions}
EK was responsible for the development and proofs of the reductive Cartan decomposition, reductive KHK decomposition and quantum assisted decomposition. OA and BNB proved the ordering of the decomposed $\kalg$-sets. OA was responsible for all the simulations, including numerical analysis, hardware runs, data analysis, and the primary writing of the manuscript. All authors contributed to the writing and editing of the manuscript.

\begin{acknowledgments}
We acknowledge helpful discussions with Arvin Kushwaha. OA, EK, and AFK acknowledge support from the National
Science Foundation under award No. 2325080: PIF: Software-Tailored Architecture for Quantum Co-Design. EK was supported by the U.S. Department of Energy (DOE) under Contract No.~DE-AC02-05CH11231 through the Office of Advanced Scientific Computing Research  Accelerated Research for Quantum Computing  Program.
BNB was supported by the U.S. Department of Energy, Advanced Scientific Computing Research, under contract number DE-SC0025384.
\end{acknowledgments}

\clearpage

\appendix
\crefalias{section}{appendix}
\renewcommand\thefigure{\thesection\arabic{figure}}  

\section{Reductive KHK Decomposition}\label{sec:reductive_KHK}
\setcounter{figure}{0}

We provide a proof for \cref{thm:reductive_KHK}. Recall from \cref{lem:reductivecd} that $\galg_{(r)} = \kalg_{(r)} \oplus \malg_{(r)}$ is a Cartan decomposition. Assume a set of coordinates $\vec{\theta}$ in a chart of $\mathcal{K}_{(r)} = \exp(\kalg_{(r)})$, and define the following function:
\begin{align}
     f_{r}(\vec{\theta}) = \langle \Phi^r(\vec{\theta}) \tilde{b}_r \Phi^r(\vec{\theta})^\dagger, \ham_{r} \rangle,
\end{align}
where $\Phi^r(\vec{\theta}) \in \mathcal{K}_{(r)}$.
Since $iH_r \in \malg_{(r)}$ and $\balg$ is Abelian, \textit{i.e.}, $i\tilde{b}_r \in \malg_{(r)}$, then \cref{thm:earppachos} tells us that if $\Phi^r_c = \Phi^r(\vec{\theta}_c)$ is an extremum of $f_r(\vec{\theta})$, then
\begin{align}
    [\Phi^{r\dagger}_c \ham_r \Phi^r_c, \tilde{b}_r] = 0.
\end{align}
The last thing to check is that $[\Phi^{r\dagger}_c \ham_r \Phi^r_c, \tilde{b}_j] = 0$ for all $j<r$. We know that $\Phi^r_c \in \mathcal{K}_{(r)}$ and $i\ham_r \in \malg_{(r)}$. Therefore, $i\Phi^{r\dagger}_c \ham_r \Phi^r_c \in \malg_{(r)}$ as well. This means that $\Phi^{r\dagger}_c \ham_r \Phi^r_c$ commutes with $\tilde{b}_j$ for all $j<r+1$, i.e., $i\Phi^{r\dagger}_c \ham_r \Phi^r_c \in \malg_{(r+1)}$.

To reduce the number of parameters, consider the coordinate patch in $\mathcal{K}_{(r)}$:
\begin{align}\label{eq:patch}
    \Phi^r(\vec{\theta}) = \Phi^r(\vec{\alpha}, \vec{\beta}) = K^r(\vec{\alpha})S^r(\vec{\beta}),
\end{align}
where $S^r(\vec{\beta})$ is a coordinate patch in the Lie subgroup $\mathcal{K}_{(r+1)} = \exp(\kalg_{(r+1)}) \subseteq \mathcal{K}_{(r)}$. This means that $S^r(\vec{\beta})$ commutes with $\tilde{b}_r$, and we recover \cref{eq:new_cost}:
\begin{align}
    \langle \Phi^r(\vec{\theta}) \tilde{b}_r \Phi^r( \vec{\theta})^\dagger, \ham_{r} \rangle = \langle K^r(\vec{\alpha}) \tilde{b}_r K^r(\vec{\alpha})^\dagger, \ham_{r} \rangle := f_r(\vec{\alpha}).
\end{align}
Since $\Phi^r(\vec{\alpha},\vec{\beta})$ is a coordinate patch in $\mathcal{K}_{(r)}$, then $|\vec{\alpha}| + |\vec{\beta}| = |\tilde{\kalg}_{(r)}|$. Similarly, we have $|\vec{\beta}| = |\tilde{\kalg}_{(r+1)}|$. We thus conclude that \cref{eq:patch} allows us to solve a $(|\tilde{\kalg}_{(r)}| - |\tilde{\kalg}_{(r+1)}|)$-dimensional optimization problem instead of a $|\tilde{\kalg}_{(r)}|$-dimensional one. \hfill$\blacksquare$

Having established that any coordinate patch of the form given in \cref{eq:patch} leads to a reductive KHK decomposition with $|\tilde{\kalg}_{(r)}| - |\tilde{\kalg}_{(r+1)}|$ angles to optimize at the $r^{\text{th}}$ step, we present two useful examples.

\subsection*{Product ansatz}

The product ansatz given by \cref{eq:prodkr} is a convenient choice of a coordinate patch because it can be easily compiled into a quantum circuit. It was shown in Apps. D and E in the Supplemental Material of Ref.~\onlinecite{kokcu2022fixed} that
\begin{align}
    \Phi^r(\vec{\theta}) = \prod_{\tilde{k}_j \in \tilde{\kalg}_{(r)}} e^{i\theta_j \tilde{k}_j}
\end{align}
is a valid choice for a KHK decomposition. We are free to order the terms in the product, so we can group them by whether or not each $\tilde{k}_j$ commutes with $\tilde{b}_r$ as follows:
\begin{align}
    \Phi^r(\vec{\alpha}, \vec{\beta}) = \underbrace{\prod_{\tilde{k}_j \in \tilde{\kalg}^r_{1\dots r-1}} e^{i\alpha_j \tilde{k}_j}}_{K^r(\vec{\alpha})}\,
    \underbrace{\prod_{\tilde{k}_j \in \tilde{\kalg}_{(r+1)}} e^{i\beta_j \tilde{k}_j}}_{S^r(\vec{\beta})}. 
\end{align}
The product on the right commutes with $b_r$, leading us to \cref{eq:prodkr}.

This ansatz can be systematically constructed directly from the dynamical Lie algebra generated by the Hamiltonian, making it convenient for generic use.

\subsection*{Compressed TFIM/TFXY ansatz}
For TFIM/TFXY, we can use another ansatz that reduces the CNOT count. Recall that for $l$ sites, we have
\begin{align}
    \kalg &= \mathrm{span}_{i\mathbb{R}}\{\widehat{X_iY_j}, \widehat{Y_iX_j}\: |\: i,j=1,2,\dots,l,\ i<j \}, \nonumber\\
    \halg &= \mathrm{span}_{i\mathbb{R}}\{Z_k \:|\: k=1,2,\dots,l\},
\end{align}
which leads to
\begin{align}
    \kalg_{(r)} = \mathrm{span}_{i\mathbb{R}}\{\widehat{X_iY_j}, \widehat{Y_iX_j}\: |\: i,j=r,r+1,\dots,l,\ i<j\}.
\end{align}
Let $D_{p,q}(\theta_j^k) := e^{i\theta^k_{1,j} \widehat{X_pY_q} + i\theta^k_{2,j} \widehat{Y_pX_q}}$ be a doublet connecting the qubits $p$ and $q$ (cf. the arrow doublets in \cref{fig:TFIMcirc}). Then for $\Phi^r(\vec{\theta})$, we can write the product ansatz:
\begin{align*}
    \Phi^r(\vec{\theta}) &= \prod_{k=r}^{l-1} \left(\prod_{j=0}^{l-k-1} D_{k,l-j}(\theta^k_j)\right),
\end{align*}
or equivalently,
\begin{align}\label{eq:tfimuncompressed}
    \Phi^r(\vec{\alpha},\vec{\beta}) = \underbrace{\left(\prod_{j=0}^{l-r-1} D_{r,l-j}(\alpha^r_j)\right)}_{K^r(\vec{\alpha})} \underbrace{\Phi^{r+1}(\vec{\beta})}_{S^r(\vec{\beta})},
\end{align}
which leads to the construction shown in \cref{fig:TFIMcirc}.

It was shown in Ref.~\onlinecite{kokcu2022fixed} that \cref{eq:tfimuncompressed} is equivalent to another ansatz given by
\begin{align*}
    \Phi^r(\vec{\theta}) &= \prod_{k=r}^{l-1} \left(\prod_{j=l}^{k+1} D_{j-1,j}(\theta^k_j)\right),
\end{align*}
which is equivalent to
\begin{align}\label{eq:tfimcompressed}
    \Phi^r(\vec{\alpha},\vec{\beta}) = \underbrace{\left(\prod_{j=l}^{r+1} D_{j-1,j}(\alpha^r_j)\right)}_{K^r(\vec{\alpha})} \underbrace{\Phi^{r+1}(\vec{\beta})}_{S^r(\vec{\beta})}.
\end{align}
\begin{figure}
    \centering
    \includegraphics[width=\columnwidth]{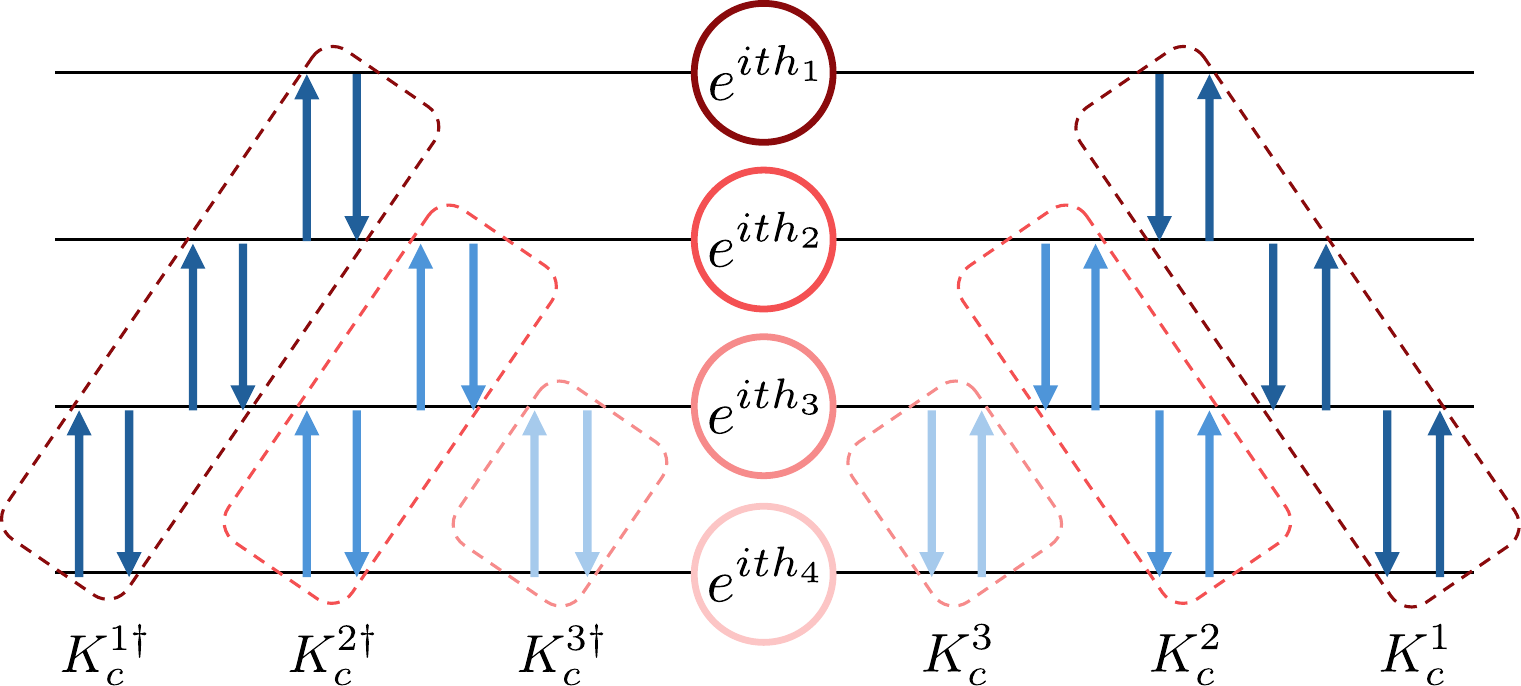}
    \caption{Compressed circuit for a 4-site TFIM time evolution unitary. An arrow doublet connected two qubits $p$ and $p+1$ depicts the unitary $e^{i\alpha X_pY_{p+1} + i\beta Y_p X_{p+1}}$. Flipping the doublet represents Hermitian conjugation. The color of a dashed box indicates the ansatz used in the reductive KHK decomposition.}
    \label{fig:compressedTFIMcirc}
\end{figure}

\noindent \cref{eq:tfimcompressed} leads to the circuit shown in \cref{fig:compressedTFIMcirc}. Note that, unlike $K^r(\vec{\alpha})$ obtained from \cref{eq:tfimuncompressed}, some terms that appear in $K^r(\vec{\alpha})$ obtained from \cref{eq:tfimcompressed} now commute with $\tilde{h}_r$. This means that the ordering of the terms in \cref{eq:tfimcompressed} matters if we want $\Phi^r(\vec{\theta})$ to be a coordinate patch in $\mathcal{K}_{(r)}$.

\section{Associative Structure of the Cartan Subalgebra}\label{sec:associative}
\setcounter{figure}{0}

When fragmenting the subalgebra $\kalg$, some sets turn out to be empty. The existence of some of these empty sets is related to the structure of the Cartan subalgebra $\halg$. 
For example, if we have $\tilde{h}_1,\tilde{h}_2,\tilde{h}_3 \in \halg$ such that $\tilde{h}_3=\tilde{h}_1\tilde{h}_2$, we know that the sets $\tilde{\kalg}^i_{jk}$ containing the elements that commute with $\tilde{h}_j$ and $\tilde{h}_k$ and anticommute with $\tilde{h}_i$ are empty for any permutation $\{i,j,k\}$ of $\{1,2,3\}$, because a Pauli string cannot commute with two of $\tilde{h}_1,\tilde{h}_2,\tilde{h}_3$ but anticommute with the third. In this section, we present a formal treatment for dealing with a Cartan subalgebra that has an additional associative structure.
\begin{widetext}
\begin{adefinition}
    For any Abelian Lie algebra $\balg = \mathrm{span}_{i\mathbb{R}}\{ \tilde{b}_j\}$,
    denote its closure under multiplication as
    \begin{align}
        \mathfrak{M}(\balg) := \mathrm{span}_{i\mathbb{R}}\{ \tilde{b}_{j_1}\tilde{b}_{j_2}\dots \tilde{b}_{j_r} \,|\, r = 1,2,...,n, \, 1\le j_1<\cdots<j_r\le n\}. 
    \end{align}
\end{adefinition}

Let us clarify what this means with an example. Take
\begin{align}
    \balg = \mathrm{span}_{i\mathbb{R}} \{Z_1,Z_2,Z_3\}.
\end{align}
Then its closure under multiplication can be found as 
\begin{align}
    \mathfrak{M}(\balg) = \mathrm{span}_{i\mathbb{R}} \{Z_1,Z_2,Z_3, Z_1Z_2,Z_1Z_3,Z_2Z_3,Z_1Z_2Z_3\},
\end{align}
which has a basis consisting of all the elements that can be produced from products of basis elements of $\balg$.
\end{widetext}
\begin{atheorem}
    Let $\balg=\mathrm{span}_{i\mathbb{R}}\{\tilde{b}_j\} \subseteq \malg$ be an Abelian Lie algebra such that $\halg \subseteq \mathfrak{M}(\balg)$ and
    \begin{align}\label{eq:new_v}
        u = \sum_{j=1}^{|\tilde{\balg}|} \gamma_j \tilde{b}_j,
    \end{align}
    where $\{\gamma_j\}$ are mutually irrational. For any $m \in \malg$, if $[m,u]=0$, then $m \in \halg$.
\end{atheorem}
\textbf{Proof} The set of group elements $e^{su}$ for $s\in\mathbb{R}$ is dense in $\exp(\mathfrak{b})$. Hence, $[m,u]=0$ implies that $[m,b] = 0$ for any $b \in \mathfrak{b}$. This in turn implies that $[m,M]=0$ for any $M \in \mathfrak{M}(\mathfrak{b})$, which also implies that $[m,h]=0$ for any $h \in \mathfrak{h}$. By \cref{thm:earppachos}, we conclude that $m \in \mathfrak{h}$.\hfill$\blacksquare$\\

We now state the main corollary:
\begin{acorollary}\label{cor:reducedkhk}
    Assume a set of coordinates $\Vec{\theta}$ in a chart of the Lie group $\mathcal{K} = \exp(\kalg)$.
    For $i\ham \in \malg$ and $K(\vec{\theta}) \in \mathcal{K}$, define the following function:
    \begin{align}\label{eq:Earp_function_app}
        f(\Vec{\theta}) = i\langle K(\Vec{\theta}) u K(\Vec{\theta})^\dagger, \ham \rangle,
    \end{align}
    where $\langle .,. \rangle$ denotes an invariant non-degenerate bilinear form on $\galg$, and $u \in \mathfrak{b}$ is an element such that $e^{su}$ for $s \in \mathbb{R}$ are dense in $\exp(\mathfrak{b})$. Suppose that $\mathfrak{h} \subseteq \mathfrak{M}(\mathfrak{b})$. Then for any {\em local extremum} of $f(\Vec{\theta})$ denoted by $\Vec{\theta}_c$, and defining the critical group element $K_c = K(\Vec{\theta}_c)$, we have
    \begin{align}
        iK(\Vec{\theta}_c)^\dagger \ham K(\Vec{\theta}_c)= iK_c^\dagger \ham K_c  \in \halg.
        \label{eq:critical_k_app}
    \end{align}
\end{acorollary}

\cref{cor:reducedkhk} allows us to perform the optimization using the reduced space $\balg$ instead of the full $\halg$. Since $\balg$ no longer has an associative structure, we will not have the issue of having a Pauli string that is a product of other Pauli strings. This naturally eliminates the empty sets that arise due to the associative structure.

It is important to note here that while the presence of an associative structure to $\halg$ is guaranteed to yield some empty sets, we can have empty sets even if there is no associative structure. For example, the TFXY algebra always has one empty set at the end, even though $\halg=\mathrm{span}_{i\mathbb{R}}\{Z_1,Z_2,\dots\}$. More precisely, working with a subalgebra that does not have an associative structure does not guarantee the elimination of all the empty sets. Nevertheless, we can make precise statements about the nature of the fragmentation of the subalgebra $\kalg$. We devote the next section to discussing the details.

Another fact worth mentioning pertains to the maximum dimension allowed for $\balg$ for any Hamiltonian $\ham$. We provide the following theorem:
\begin{atheorem}
    Consider a Hamiltonian $\ham$ that generates a DLA $\galg(\ham)\subseteq \mathfrak{su}(2^n)$ with a CD $\galg=\kalg\oplus\malg$ such that $i\ham\in\malg$, and a Cartan subalgebra $\halg \subseteq \malg$. Then $\halg \subseteq \mathfrak{M}(\balg)$ for some Abelian subalgebra $\balg \subseteq \mathfrak{su}(2^n)$ 
    with $\dim{\balg} \leq n$.
\end{atheorem}
\textbf{Proof} 
The Cartan subalgebra of $\mathfrak{su}(2^n)$ can be chosen as $\mathfrak{M}(\balg)$, where $\balg=\mathrm{span}_{i\mathbb{R}}\{Z_1,Z_2,\dots,Z_n\}$. Recall that the Cartan subalgebra is a maximal Abelian subalgebra, and all Cartan subalgebras are conjugate. Since $\halg$ is Abelian, it can be conjugated to a subspace of $\mathfrak{M}(\balg)$.
\hfill$\blacksquare$\\

This theorem implies that at most we need to solve $n$ optimization problems.

\section{Fragmentation of the $\kalg$ Subalgebra}\label{sec:ksizes}
\setcounter{figure}{0}

In this section, we show that there is no preferred sequence of doing the reductive KHK decomposition. We start by making two definitions:
\begin{adefinition}
    An \emph{induced subgraph} of a graph $G$ is a graph whose vertex set is a subset of the vertices of $G$ and whose edges are all the edges in $G$ that have both endpoints included in that subset.
\end{adefinition}
\begin{adefinition}
    An \emph{induced path} is a path that is an induced subgraph.
\end{adefinition}

We can then make the following statement about the frustration graph of a DLA $\galg$:
\begin{atheorem}\label{thm:path}
    Consider a DLA $\galg$ and two commuting Pauli strings $b_1,b_2 \in \galg$. If $b_1$ and $b_2$ belong to the same connected component of the frustration graph of $\galg$, then the distance between $b_1$ and $b_2$ is $2$, i.e., there exists at least one Pauli string $g\in \galg$ such that $[g,b_1]\neq 0$ and $[g,b_2] \neq 0$.
\end{atheorem}
\textbf{Proof} Consider any sequence $g_0,g_1,g_2,\dots, g_{n+1}$ that forms an induced path from $g_0 = b_1$ to $g_{n+1} = b_2$. Hence for $0<i<n+1$, $[g_i,g_j]\neq 0$ for $j=i\pm 1$ and $[g_i,g_j]=0$ otherwise. This implies that $\big[[g_i,g_{i+1}],g_{i+2}\big]\neq 0$. Inductively, we conclude that $g=g_1\dots g_n$ is in $\galg$. We see that $[g,b_1]\neq 0$ and $[g,b_2]\neq 0$.\hfill$\blacksquare$\\

\begin{figure}[t]
    \centering
    \includegraphics[width=0.4\columnwidth]{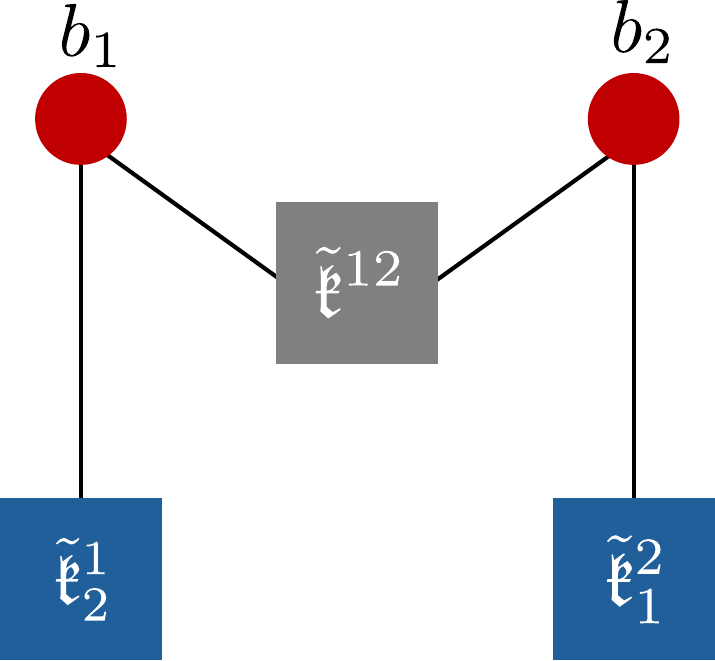}
    \caption{Graphical illustration of \cref{lem:twoindex}. The red dots represent the two commuting elements $b_1$ and $b_2$. Boxes represent the different $\tilde{\kalg}$ sets. The lines connect the sets with the element with which they anticommute. If $\tilde{\kalg}^{12}$ (gray) is non-empty, then $|\tilde{\kalg}^1_2| = |\tilde{\kalg}^2_1|$ (blue). Note that this is not a full frustration graph; only lines attached to the red dots are drawn.}
    \label{fig:2sets}
\end{figure}

If we associate a CD to $\galg$, then we immediately get an additional result:
\begin{acorollary}\label{cor:kexist}
    In addition to the assumptions of \cref{thm:path}, suppose that $\galg = \kalg \oplus \malg$ is a Cartan decomposition such that $b_1,b_2 \in \malg$. Then there exists at least one Pauli string $k\in \kalg$ such that $[k,b_1]\neq 0$ and $[k,b_2]\neq 0$, i.e., $\kalg^{12}$ is not empty.
\end{acorollary}
\textbf{Proof} By \cref{thm:path}, there is a Pauli string $g\in \galg$ that does not commute with both $b_1$ and $b_2$. If $g\in \kalg$, we are done. If $g\in \malg$, we can construct the element $g' = gb_1$. Since $[\malg,\malg]\subseteq \kalg$, then $g' \in \kalg$.\hfill$\blacksquare$\\

Now we are ready to start finding some relations between the different fragmented $\kalg$ subspaces. With \cref{def:ksets} in mind, we derive the following lemma.

\begin{alemma}\label{lem:twoindex}
    Given a DLA $\galg$ and its Cartan decomposition $\galg = \kalg \oplus \malg$, take two commuting Pauli strings $b_1,b_2 \in \malg$ that belong to the same connected component of the frustration graph of $\galg$. Then $|\tilde{\kalg}^1_2| = |\tilde{\kalg}^2_1|$.
\end{alemma}
\textbf{Proof} From \cref{cor:kexist}, we know that $\tilde{\kalg}^{12}$ is not empty. Without loss of generality, assume that $|\tilde{\kalg}^1_2| > |\tilde{\kalg}^2_1|$. Pick any element $d\in \tilde{\kalg}^{12}$. Divide $\tilde{\kalg}^1_2$ into two disjoint subsets $\tilde{\kalg}_c$ and $\tilde{\kalg}_a$ such that $[d,\tilde{\kalg}_c]=\{0\}$ and $[d,\tilde{\kalg}_a]\neq \{0\}$. Next construct the two sets $\mathfrak{p}_c = b_2db_1\tilde{\kalg}_c$ and $\mathfrak{p}_a=d\tilde{\kalg}_a$. Observe that $\mathfrak{p}_c$ and $\mathfrak{p}_a$ are also disjoint since $[d,\mathfrak{p}_c]=\{0\}$ while $[d,\mathfrak{p}_a]\neq \{0\}$. This construction also puts $\mathfrak{p}_c$ and $\mathfrak{p}_a$ in $\tilde{\kalg}^2_1$, \textit{i.e.}, $\mathfrak{p}_c \cup \mathfrak{p}_a \subseteq \tilde{\kalg}^2_1$. Finally, note that $|\mathfrak{p}_c|=|\tilde{\kalg}_c|$ and $|\mathfrak{p}_a| = |\tilde{\kalg}_a|$. Since $\mathfrak{p}_c$ and $\mathfrak{p}_a$ are disjoint, we have $|\mathfrak{p}_c \cup \mathfrak{p}_a| = |\mathfrak{p}_c| + |\mathfrak{p}_a|=|\tilde{\kalg}^1_2|$. But $|\mathfrak{p}_c \cup \mathfrak{p}_a| \leq |\tilde{\kalg}^2_1|$, which is a contradiction. We thus conclude that $|\tilde{\kalg}^1_2| = |\tilde{\kalg}^2_1|$.\hfill$\blacksquare$

\begin{acorollary}\label{cor:1indexsym}
Under the above assumptions, we have $|\tilde{\kalg}^1| = |\tilde{\kalg}^2|$.
\end{acorollary}

\begin{figure}[t]
    \centering
    \includegraphics[width=\columnwidth]{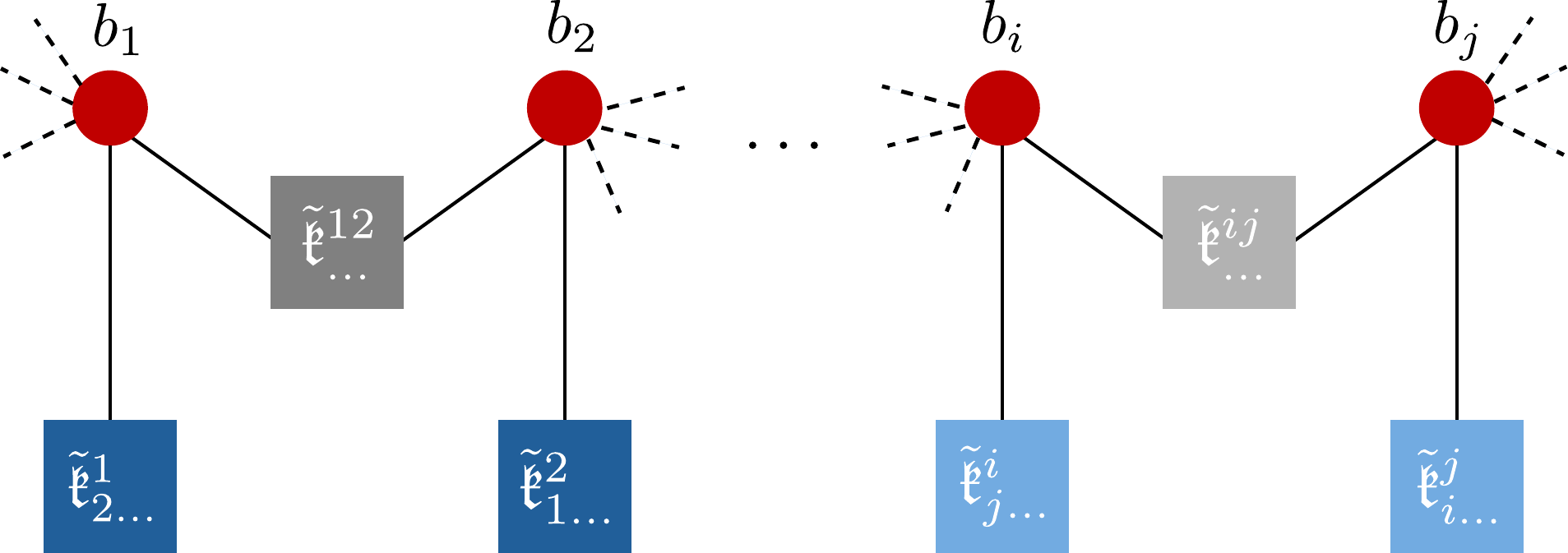}
    \caption{Graphical illustration of \cref{cor:specialtwoindex}. For relevance, only $\tilde{\kalg}$ sets attached to one and two lines are shown. If $\tilde{\kalg}^{12}_{\dots}$ (dark gray) is non-empty, then $|\tilde{\kalg}^1_{2\dots}| = |\tilde{\kalg}^2_{1\dots}|$ (dark blue). Similarly if $\tilde{\kalg}^{ij}_{\dots}$ (light gray) is non-empty, then $|\tilde{\kalg}^i_{j\dots}| = |\tilde{\kalg}^j_{i\dots}|$ (light blue).}
    \label{fig:moresets}
\end{figure}

\cref{fig:2sets} illustrates \cref{lem:twoindex}. In general, if we have more than two commuting strings, \cref{cor:1indexsym} tells us that $|\tilde{\kalg}^i|$ is the same for any choice of index. We also have the following corollary.
\begin{acorollary}\label{cor:specialtwoindex}
    If $\tilde{\kalg}^{ij}_{r_1r_2\dots}$ is non-empty, then $|\tilde{\kalg}^i_{jr_1r_2\dots}| = |\tilde{\kalg}^j_{ir_1r_2\dots}|$.
\end{acorollary}
The proof is similar to \cref{lem:twoindex}'s. See \cref{fig:moresets} for a graphical representation of \cref{cor:specialtwoindex}.

The structure of the frustration graph can give us some insight on the associative structure of the Abelian subalgebra picked. We provide the following theorem:
\begin{atheorem}\label{thm:123impliesassociative}
    Given a DLA $\galg$ and its Cartan decomposition $\galg = \kalg \oplus \malg$, take three commuting Pauli strings $b_1,b_2,b_3 \in \malg$. Then $\tilde{\kalg}^{123}$ is non-empty only if $b_1b_2b_3 \in \malg$.
\end{atheorem}
\textbf{Proof} Suppose that $\tilde{\kalg}^{123}$ is not empty and pick any $d\in \tilde{\kalg}^{123}$. Then $db_1 \in \malg$, $db_1b_2 \in \kalg$, $db_1b_2b_3 \in \malg$, and finally $db_1b_2b_3d = b_1b_2b_3 \in \malg$. Therefore if $b_1b_2b_3 \notin \malg$, then $\tilde{\kalg}^{123}$ must be empty.\hfill$\blacksquare$\\

The absence of an associative structure to an Abelian set that belongs to the same component of the frustration graph thus implies that the only non-empty sets that connect the elements are of the form $\tilde{\kalg}^{r_1r_2}_{r_3r_4\dots}$. This is enough to prove \cref{thm:ksizes} in this case. The general case requires some more work. We provide two additional lemmas that we need in order to prove \cref{thm:ksizes}, and we mention the special case where an associative structure is absent at the end of the section.
\begin{alemma}\label{lem:3nonempty}
    Given a DLA $\galg$ and its Cartan decomposition $\galg = \kalg \oplus \malg$, take an Abelian subalgebra $\balg \subseteq \malg$ such that its Pauli string basis elements $b_1,b_2,\dots, b_n$ belong to the same connected component of the frustration graph of $\galg$ and $\tilde{\kalg}^{r-1}_{1\dots r-2}$ is non-empty for some $r\leq n$. Then $\tilde{\kalg}^r_{1\dots r-1}$ is non-empty only if $\tilde{\kalg}^{r-1,r}_{1\dots r-2}$ is non-empty.
\end{alemma}
\textbf{Proof} Suppose that $\tilde{\kalg}^{r-1}_{1\dots r-2}$ and $\tilde{\kalg}^r_{1\dots r-1}$ are non-empty but $\tilde{\kalg}^{r-1,r}_{1\dots r-2}$ is empty. Since according to \cref{cor:kexist}, $\tilde{\kalg}^{r-1,r}$ is non-empty, we get that $\tilde{\kalg}^{s,r-1,r}$ must be non-empty for some $s<r-1$. \cref{thm:123impliesassociative} then tells us that $d=b_sb_{r-1}b_r \in \malg$. 

\begin{figure}[t]
    \centering
    \includegraphics[width=0.8\columnwidth]{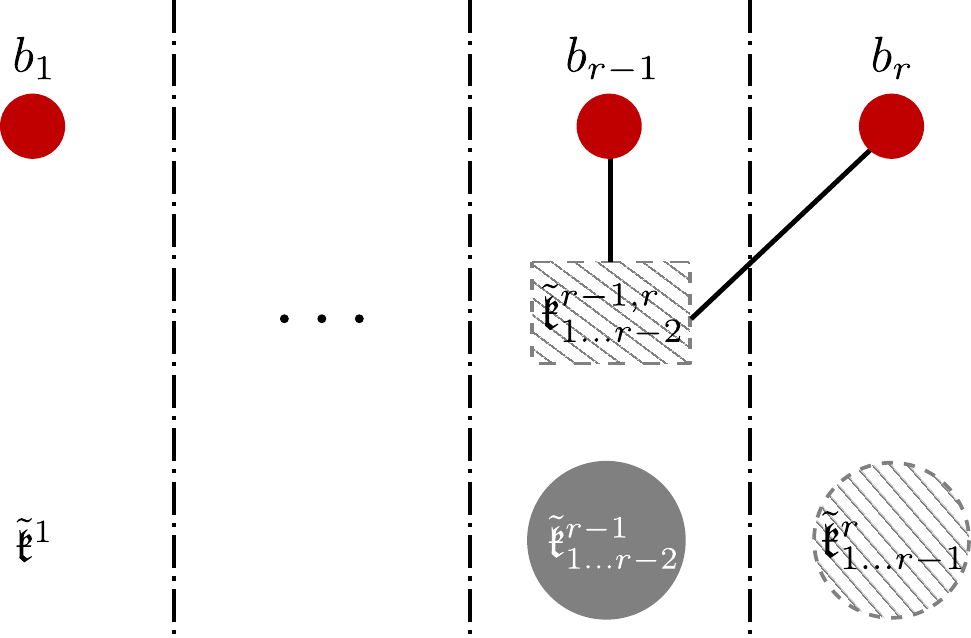}
    \caption{Graphical illustration of \cref{lem:3nonempty}. Each region (labeled at the bottom) separated by the vertical dividers containing a red dot $b_i$ contains all the elements in $\tilde{\kalg}^i_{1\dots i-1}$ that anticommute with $b_i$ and commute with $b_j$ for $j<i$. If $\tilde{\kalg}^{r-1}_{1\dots r-2}$ (solid gray circle) is non-empty, then for $\tilde{\kalg}^r_{1\dots r-1}$ (striped gray circle) to be non-empty, $\tilde{\kalg}^{r-1,r}_{1\dots r-2}$ must be non-empty as well. Lines attached to $\tilde{\kalg}^{r-1,r}_{1\dots r-2}$ are shown to indicate that this set anticommutes with $b_r$ as well as $b_{r-1}$.}
    \label{fig:next_set_nonempty}
\end{figure}

\cref{fig:next_set_nonempty} illustrates \cref{lem:3nonempty}. Next, note that the set of elements in $\tilde{\kalg}^{r-1}_{1\dots r-2}$ that also commute with $d$ is simply the set $\tilde{\kalg}^{r-1,r}_{1\dots r-2}$, which we assumed empty. Therefore, all the elements in the non-empty set $\tilde{\kalg}^{r-1}_{1\dots r-2}$ anticommute with $d$. This allows us to use \cref{cor:specialtwoindex} to conclude that the size of the subset of $\tilde{\kalg}^{r-1}_{1\dots r-2}$ that also commutes with $d$ ---which is empty under our assumptions--- is equal to the size of the subset of $\tilde{\kalg}_{1\dots r-1}$ that anticommutes with $d$. Finally, note that this subset is equal to $\tilde{\kalg}^r_{1\dots r-1}$. Therefore the only way for $\tilde{\kalg}^r_{1\dots r-1}$ to be non-empty is if $\tilde{\kalg}^{r-1,r}_{1\dots r-2}$ is not empty.\hfill$\blacksquare$

\begin{acorollary}\label{cor:nonemptysym}
    In addition to the assumptions of \cref{lem:3nonempty}, if $\tilde{\kalg}^r_{1\dots r-2}$ is also non-empty, then $|\tilde{\kalg}^{r-1}_{1\dots r-2}| = |\tilde{\kalg}^r_{1\dots r-2}|$. Equivalently, $|\tilde{\kalg}^{r-1}_{1\dots r-2, r}|=|\tilde{\kalg}^r_{1\dots r-1}|$.
\end{acorollary}
\textbf{Proof} Consider $\tilde{\kalg}^{r-1}_{1\dots r-2} = \tilde{\kalg}^{r-1,r}_{1\dots r-2}\cup \tilde{\kalg}^{r-1}_{1\dots r-2,r}$ and $\tilde{\kalg}^{r}_{1\dots r-2} = \tilde{\kalg}^{r-1,r}_{1\dots r-2}\cup \tilde{\kalg}^{r}_{1\dots r-1}$. If $\tilde{\kalg}^r_{1\dots r-1}$ is empty, then $\tilde{\kalg}^{r-1,r}_{1\dots r-2}$ has to be non-empty, and consequently $|\tilde{\kalg}^{r-1}_{1\dots r-2,r}|=|\tilde{\kalg}^r_{1\dots r-1}|=0$, which leads to $|\tilde{\kalg}^{r-1}_{1\dots r-2}| = |\tilde{\kalg}^r_{1\dots r-2}|$. If instead $\tilde{\kalg}^r_{1\dots r-1}$ is non-empty, then \cref{lem:3nonempty} tells us that $\tilde{\kalg}^{r-1,r}_{1\dots r-2}$ is non-empty, and we again arrive at $|\tilde{\kalg}^{r-1}_{1\dots r-2}| = |\tilde{\kalg}^r_{1\dots r-2}|$. \hfill $\blacksquare$

\begin{acorollary}\label{cor:emptysym}
    If $\tilde{\kalg}^r_{1\dots r-2}$ is empty, then $\tilde{\kalg}^{r-1,r}_{1\dots r-2}$ and $\tilde{\kalg}^r_{1\dots r-1}$ are also empty. Then $\tilde{\kalg}^{r-1}_{1\dots r-2,r} = \tilde{\kalg}^{r-1}_{1\dots r-2}$.
\end{acorollary}

\begin{figure}[t]
    \centering
    \includegraphics[width=0.8\columnwidth]{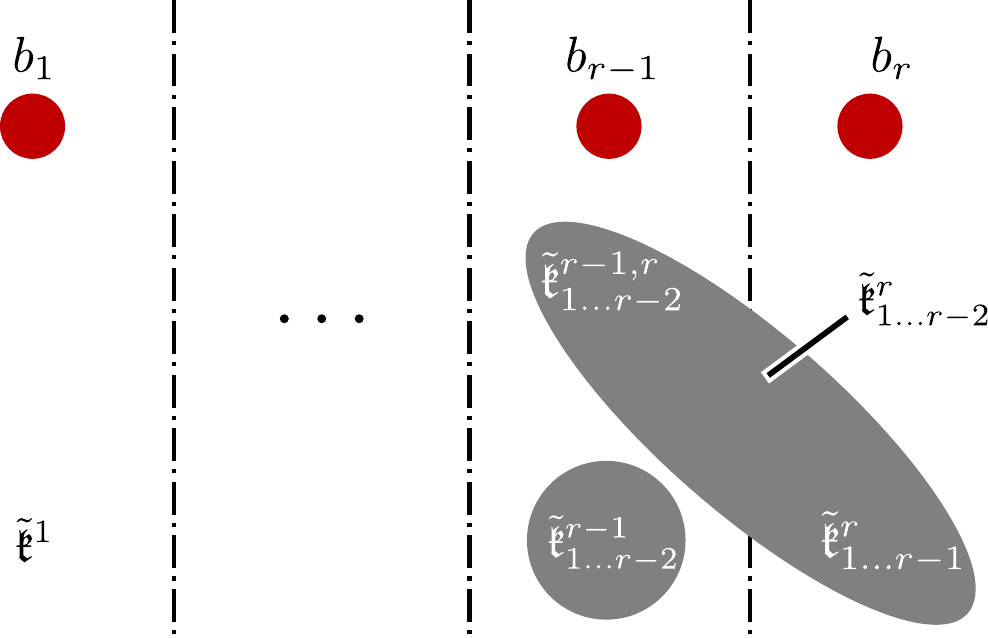}
    \caption{Graphical illustration of \cref{cor:nonemptysym}. The setup is similar to \cref{fig:next_set_nonempty}. If $\tilde{\kalg}^{r-1}_{1\dots r-2}$ and $\tilde{\kalg}^r_{1\dots r-2}$ are both non-empty, then they must be equal in size.}
    \label{fig:reordering_equality}
\end{figure}

Although \cref{cor:emptysym} is obvious, we include it for completeness. Together with \cref{cor:nonemptysym}, we conclude that whichever element we choose next in the sequence does not alter the size of the optimization problem. If we pick an empty set, then it can be discarded, \textit{i.e.}, moved to the end of the sequence. This proves the index symmetry of the sets. All that is left is to show that the sequence is strictly decreasing until we run out of non-empty sets. This follows directly from \cref{cor:nonemptysym} (see \cref{fig:reordering_equality} for a graphical representation).

\begin{alemma}\label{lem:2inequality}
    Given a DLA $\galg$ and its Cartan decomposition $\galg = \kalg \oplus \malg$, take an Abelian subalgebra $\balg \subseteq \malg$ such that its Pauli string basis elements $b_1,b_2,\dots b_n$ belong to the same connected component of the frustration graph of $\galg$. If $\tilde{\kalg}^{r-1}_{1\dots r-2}$ and $\tilde{\kalg}^r_{1\dots r-1}$ are non-empty for some $r\leq n$, then $|\tilde{\kalg}^{r-1}_{1\dots r-2}| > |\tilde{\kalg}^r_{1\dots r-1}|$.
\end{alemma}
\textbf{Proof} From \cref{cor:nonemptysym} we have $|\tilde{\kalg}^{r-1}_{1\dots r-2}| = |\tilde{\kalg}^r_{1\dots r-2}|=|\tilde{\kalg}^{r-1,r}_{1\dots r-2}|+|\tilde{\kalg}^{r}_{1\dots r-1}|$. \cref{lem:3nonempty} tells us that $\tilde{\kalg}^{r-1,r}_{1\dots r-2}$ is non-empty. We thus conclude that $|\tilde{\kalg}^{r-1}_{1\dots r-2}|>|\tilde{\kalg}^r_{1\dots r-1}|$.\hfill$\blacksquare$\\

Now we can derive the proof of \cref{thm:ksizes} from \cref{cor:nonemptysym}, \cref{cor:emptysym} and \cref{lem:2inequality}. Here we restate \cref{thm:ksizes} for convenience.
\ksizes*

\noindent\textbf{Proof} \cref{cor:1indexsym} tells us that $|\tilde{\kalg}^i|$ is the same for any $b_i$. Without loss of generality we fix $i=1$. Then we choose the second element such that $\tilde{\kalg}^2_1$ is non-empty. By \cref{cor:nonemptysym}, the size of this set will be the same for any choice of $b_i$ so long as the set is non-empty. By \cref{lem:2inequality}, this set will be smaller than $\tilde{\kalg}^1$. We then keep going until we run out of non-empty sets. \cref{cor:nonemptysym} also tells us that we can reach any permutation of the indices by a series of interchanges of subsequent pairs. Similarly, permuting the indices of the empty sets within each other does not change anything. Finally, if we swap $b_r$ and $b_{r+1}$, then if $\tilde{\kalg}^{r+1}_{1\dots r-2}$ is non-empty, the sequence is preserved by \cref{cor:nonemptysym}. On the other hand, if $\tilde{\kalg}^{r+1}_{1\dots r-1}$ is empty, we will have introduced an empty set in the sequence. Subsequent interchanges between this empty set and the preceding non-empty sets can be treated similarly.\hfill$\blacksquare$\\

We can now return to the special case where we do not have any associative structure. In this case, due to \cref{thm:123impliesassociative}, the basis elements of the subalgebra can only be connected through the sets $\tilde{\kalg}^{r_1r_2}_{r_3r_4\dots}$. This leads to the following result:
\begin{atheorem}\label{thm:noassociative}
    Given a DLA $\galg$ and its Cartan decomposition $\galg = \kalg \oplus \malg$, take an Abelian subalgebra $\balg \subseteq \malg$  such that its Pauli string basis elements $b_1,b_2,\dots,b_n \in \balg$ belong to the same connected component of the frustration graph of $\galg$. If the basis has no associative structure, then
    \begin{align}\label{eq:linear}
        |\tilde{\kalg}^r_{1\dots{r-1}}| = |\tilde{\kalg}^1_{2\dots n}| + (n-r)|\tilde{\kalg}^{12}_{3\dots n}|,
    \end{align}
    for any $r \leq n$.
\end{atheorem}
\textbf{Proof} The basis elements are connected, so $\tilde{\kalg}^{r_1r_2}_{r_3r_4\dots}$ is non-empty for any collection of indices. This implies that $|\tilde{\kalg}^{r_1}_{r_2r_3\dots}|$ is the same for any collection of indices, per \cref{cor:specialtwoindex}. Moreover, note that $\tilde{\kalg}^{r_1r_2}_{r_3r_4\dots} = \tilde{\kalg}^{r_1}_{r_3r_4\dots} \backslash \tilde{\kalg}^{r_1}_{r_2r_3\dots}$. Since $\tilde{\kalg}^{r_1}_{r_2r_3\dots}$ is a subset of $\tilde{\kalg}^{r_1}_{r_3r_4\dots}$, we have $|\tilde{\kalg}^{r_1r_2}_{r_3r_4\dots}| = |\tilde{\kalg}^{r_1}_{r_3r_4\dots}| - |\tilde{\kalg}^{r_1}_{r_2r_3\dots}|$. The right-hand side is the same for any collection of indices, which means that $|\tilde{\kalg}^{r_1r_2}_{r_3r_4\dots}|$ also does not depend on our choice of indices.

Next, we turn to $\tilde{\kalg}^r_{1\dots r-1}$. We can write it as
\begin{align}
    \tilde{\kalg}^r_{1\dots r-1} = \Big(\bigcup_{i \neq r} \tilde{\kalg}^{ir}_{1\dots n}\Big) \cup \tilde{\kalg}^r_{1\dots n}, 
\end{align}
where the lower indices do not contain the upper indices. The big union includes $n-r$ sets. From the preceding discussion, we know that these sets all have the same size. In particular, they all have the same size as $\tilde{\kalg}^{12}_{3\dots n}$. We also know that $|\tilde{\kalg}^r_{1\dots n}| = |\tilde{\kalg}^1_{2\dots n}|$. This leads us to \cref{eq:linear}.\hfill$\blacksquare$\\

\cref{thm:noassociative} explains the linear trend we observed for the TFIM and TFXY models in \cref{fig:modelsubspaces}. Generally, any model whose Cartan subalgebra does not have any associative structure will result in a linear trend. The slope of the line would be $-|\tilde{\kalg}^{12}_{3\dots n}|$, and the $y$-intercept would be $|\tilde{\kalg}^1_{2\dots n}| + n |\tilde{\kalg}^{12}_{3\dots n}|$, where $n$ is the dimension of the Cartan subalgebra.

\section{Preparing initial states}\label{sec:statecircuits}
\setcounter{figure}{0}

\begin{figure*}[t]
    \centering
    \includegraphics[width=\textwidth]{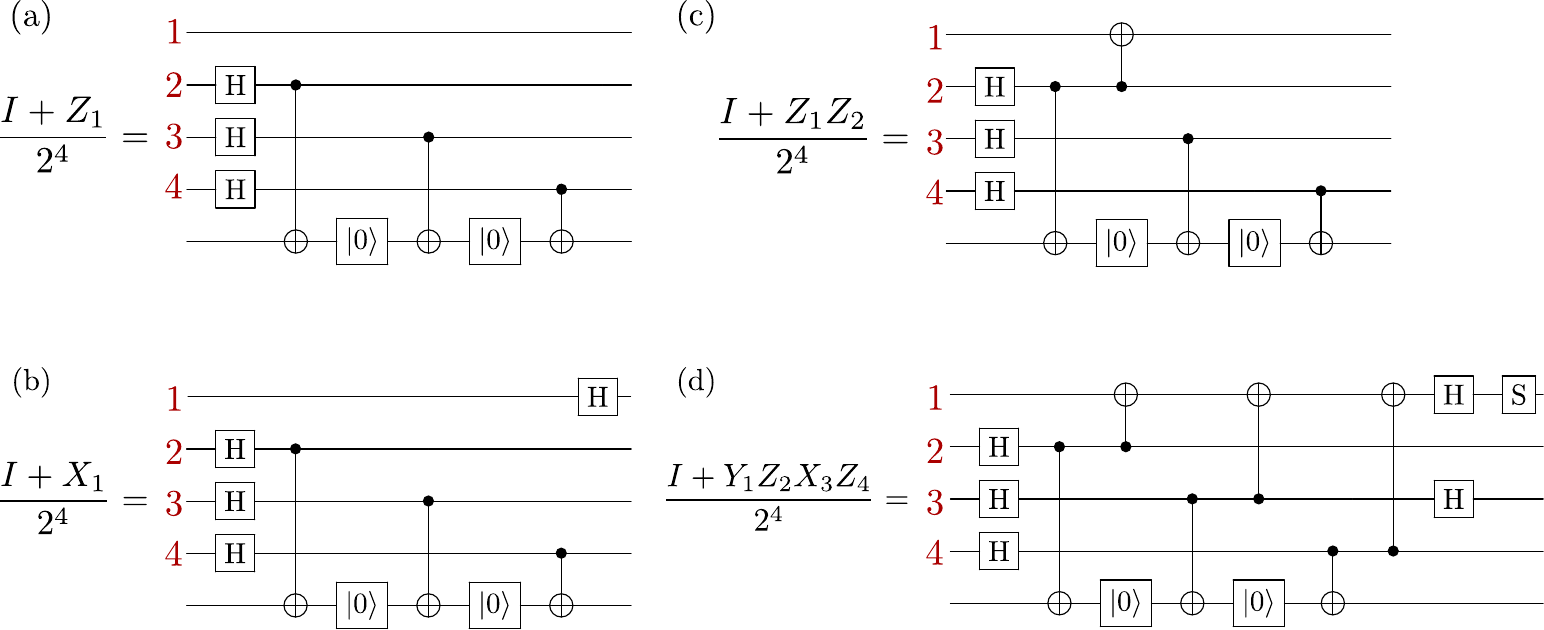}
    \caption{Different state preparations. (a) Starting point for a generic density matrix of the form $(I+P)/2^n$. To construct the state $(I+Z_i)/2^n$, we act with Hadamard (H) gates on all $j\neq i$ qubits and connect it to an ancilla qubit with a CNOT as shown, reseting after each gate. (b) Any other non-identity qubit after the first one is prepared by entangling it with that first qubit as shown. (c) If instead we want $(I+X_i)/2^n$ or $(I+Y_i)/2^n$ we act with an H gate or H and S gates on the $i$th qubit. (d) A generic state preparation procedure.}
    \label{fig:statprep}
\end{figure*}
In \cref{sec:QAD}, we recast our cost function as an expectation value of a density matrix that we can measure on a quantum computer. The density matrix we need to prepare is
\begin{align}\label{eq:thermalstate}
    \rho = \frac{I + \sigma}{2^n}, \qquad \sigma = \bigotimes_{i=1}^n P_i
\end{align}
where $I$ is the $2^n$-dimensional identity, and $P_i$ are Pauli matrices or the identity. In this section, we develop quantum circuits to prepare such density matrices with the help of $n-1$ ancilla qubits, or alternatively a single ancilla that is repeatedly reset.

We initialize the state $|0\rangle ^{\otimes n} \otimes |0\rangle_a ^{\otimes n-1}$. The second set of qubits are ancillas that are going to be traced out in the end. Next, we identify the first non-identity $P_j$ that appears in $\sigma$. For notational convenience, we reorder our qubits such that $P_j$ is the first qubit in our product. We apply Hadamard gates to all the other physical qubits:
\begin{align}
    \prod_{i = 2}^n \mathrm{H}_i |0\rangle ^{\otimes n} \otimes |0\rangle ^{\otimes n-1} = \Big(|0\rangle |+\rangle^{\otimes n - 1}\Big ) \otimes |0\rangle_a ^{\otimes n-1}.
\end{align}
We further apply CNOT gates to the same qubits, where each physical qubit is the control to one of the ancilla qubits. This gives us
\begin{align}
    &\prod_{i = 2}^n \mathrm{CNOT}_{i, a} \Big(|0\rangle |+\rangle^{\otimes n - 1}\Big ) \otimes |0\rangle_a ^{\otimes n}\nonumber \\
    = &\frac{1}{2^{(n-1)/2}}\sum_{v\in \mathbb{Z}_2^{\otimes n-1}} |0v\rangle\otimes |v\rangle_a,
\end{align}
where $v$ are all the bitstrings of length $n-1$. Now let us see what happens when we trace out the ancilla qubits. We write the state as a density matrix $|\psi\rangle \langle \psi|$. We find
\begin{align}\label{eq:proj1}
    &\frac{1}{2^{n-1}}\sum_{v,w\in \mathbb{Z}_2^{\otimes n-1}} |0v\rangle \langle 0w| \ \mathrm{Tr}\big(|v\rangle_a \langle w|_a\big)\nonumber \\
    = &\frac{1}{2^{n-1}}\sum_{v\in \mathbb{Z}_2^{\otimes n-1}} |0v\rangle \langle 0v|.
\end{align}
The partial trace thus yields the projection operator (normalized to unit trace) on the $|0\rangle$ state of the $j$th qubit. In matrix representation, we have now generated the state
\begin{align}\label{eq:I+Z}
    \rho = \frac{I + Z_j}{2^n}.
\end{align}

Next we continue with the other non-identity Pauli matrices that appear in $\sigma$. Consider again \cref{eq:proj1}. For ease of notation, reorder the qubits such that the next non-identity Pauli $P_k$ is the second qubit. We apply a CNOT gate with the $j$th qubit as the target and the $k$th qubit as the control. This gives us
\begin{align}
    &\frac{1}{2^{n-1}} \sum_{v\in \mathbb{Z}_2^{\otimes n-2}} \mathrm{CNOT}_{k,j} \Big(|00v\rangle \langle00v| + |01v\rangle \langle 01v| \Big) \nonumber \\
    = &\frac{1}{2^{n-1}} \sum_{v\in \mathbb{Z}_2^{\otimes n-2}} |00v\rangle \langle 00v| + |11v\rangle \langle 11v|,
\end{align}
which is the projection onto the the space spanned by $|00\rangle, |11\rangle$. Similar to \cref{eq:I+Z}, we have
\begin{align}
    \rho = \frac{I + Z_j Z_k}{2^{n}}.
\end{align}
In general, if we want to include $Z_i$, we apply a CNOT to the $i$th qubit and the $j$th qubit. Finally, we can use $X = \mathrm{H}Z\mathrm{H}$ and $Y = \mathrm{SH}Z\mathrm{HS}^\dagger$ to apply one qubit rotations to arrive at the desired state \cref{eq:thermalstate}. For a general Pauli string with Pauli weight $w$, a total of $n+w-2$ CNOTs are needed.

\end{document}